\documentclass{aa}  
\usepackage{graphicx}
\usepackage{txfonts}
\usepackage{textcomp}

\usepackage{booktabs} 
\usepackage{float}    
\usepackage{siunitx}  
\usepackage{makecell}
\usepackage{tabularx}
\usepackage{threeparttable}
\usepackage{graphicx}
\usepackage{amsmath}
\usepackage{adjustbox}
\usepackage{longtable}
\usepackage{geometry} 
\usepackage{array}
\usepackage{caption} 
\usepackage{subcaption} 
\usepackage{lscape} 
\usepackage{hyperref}
\usepackage{float} 

\usepackage{rotating}

\begin{document} 

   \title{STEP survey:}  
   \subtitle{III. STEPping stones between the clouds: the star formation history of the Magellanic Bridge \thanks{Based on data products from observations collected at the European Organization for Astronomical Research in the Southern Hemisphere, under ESO programs 099.D-0673(A); 098.D-0579(A); 097.D0209(A); 096.D-0535(A); 095.D-0132(A); 094.D-0492(A); 093.D0174(A); 092.D-0214(A); 091.D-0574(A); 090.D-0172(A); 089.D0258(A); and 088.D-4014(A)}}

   \titlerunning{STEP III}


   \author{
   F. Ficara\inst{1,2} \and
   V. Ripepi\inst{1} \and
   M. Cignoni\inst{3,4,5} \and
   M. Gatto\inst{1} \and
   M. Marconi\inst{1} \and
   M. Tosi\inst{4} \and
   M. Bellazzini\inst{4} \and
   E. K. Grebel\inst{6} \and
   M. R. Cioni\inst{7} \and
   C. Tortora\inst{1} \and
   A. Mercurio\inst{1,2,8}
          }

   \institute{INAF Osservatorio Astronomico di Capodimonte, Salita Moiariello 16, 80131 Napoli, Italy
   \and
   Università di Salerno, Dipartimento di Fisica “E.R. Caianiello”, Via Giovanni Paolo II 132, 84084 Fisciano (SA), Italy.
   \and
   Dipartimento di Fisica, Università di Pisa, Largo Bruno Pontecorvo 3, 56127, Pisa, Italy
   \and
   INAF- Osservatorio di Astrofisica e Scienza dello Spazio di Bologna, via Piero Gobetti 93/3, 40129 Bologna, Italy
   \and
   INFN, Largo B. Pontecorvo 3, 56127, Pisa, Italy
   \and
   Astronomisches Rechen-Institut, Zentrum f\"ur Astronomie der Universit\"at Heidelberg, M\"onchhofstra{\ss}e.~12--14, 69120 Heidelberg, Germany
   \and
   Leibniz-Institut f\"ur Astrophysik Potsdam, An der Sternwarte 16, 14482 Potsdam, Germany
   \and
   INFN–Gruppo Collegato di Salerno–Sezione di Napoli, Dipartimento di Fisica “E.R. Caianiello”, Università di Salerno, Via Giovanni Paolo II, 132, 84084 Fisciano (SA), Italy.
   }

   \date{Received XXX; accepted XXX}

 
  \abstract
   {The Magellanic Clouds (MCs) offer a unique laboratory for studying galaxy interaction and the evolution of dwarf galaxies. By investigating when and how stars formed, the star formation history (SFH) is a powerful tool to provide constraints for dynamical modeling of the system's past interactions and understand the processes of stripping and triggered star formation in tidally influenced environments.} 
   {We aim to reconstruct the SFH of the Magellanic Bridge, the gaseous and stellar stream connecting the two Clouds. We used data from the deep optical STEP survey, which covers 54 $\mathrm{deg\, {^{2}}}$ across the Small Magellanic Cloud (SMC) and the Bridge, reaching stars below the oldest main sequence turnoff at the distance of the MCs.}
   {We applied the synthetic color-magnitude diagram (CMD) technique to 14 deg$^2$ of STEP data. We constructed two libraries of synthetic stellar populations based on the PARSEC-COLIBRI and BaSTI stellar evolutionary models, with metallicities in the range $-2.0\leq[$Fe/H$]\leq0$ across the whole Hubble time.}
   {We find a clear peak of recent star formation $\sim100$ Myr ago in the Magellanic Bridge, which becomes increasingly pronounced toward the SMC. The low metallicity of this population suggests that it formed from gas stripped from the SMC during its most recent close encounter with the LMC. In the eastern part of the Bridge (LMC side), the star formation peaks at earlier times, around 10 Gyr and 2 Gyr ago. We estimate a total stellar mass in the Bridge of $ (5.1 \pm 0.2) \times 10^5 M_\odot$ and a present-day stellar metallicity of $[$Fe/H$]\sim-0.6$ dex, close to SMC value.}
   {}


   \maketitle
%

\section{Introduction}

   The SMC and Large Magellanic Cloud (LMC) are the closest irregular galaxies orbiting the Milky Way (MW), and are considered the closest known pair of interacting dwarfs. According to the Lambda Cold Dark Matter (\(\Lambda\)CDM) model, smaller structures form first and then merge to create larger galaxies \citep{1978MNRAS.183..341W, 1999ApJ...524L..19M, 2017NatAs...1E..25S}. Hence, dwarf galaxies, which reside in the smallest dark matter sub-halos, are considered optimal candidates for the role of ‘building blocks’ in galaxy formation. The MCs provide an ideal laboratory for studying the evolution of interacting dwarf galaxies, as well as likely future contributors to the MW. Their proximity allows for deep observations with ground-based telescopes, enabling the reconstruction of their spatially resolved SFHs with great detail. The LMC is a barred Magellanic spiral located at $\sim50$ kpc \citep{2019Natur.567..200P}, while its smaller companion, the SMC, is a dwarf irregular galaxy (dIrr) at $\sim62.5$ kpc \citep{2020ApJ...904...13G}. Their mutual gravitational interactions have distorted the LMC's disk \citep{2012AJ....144..106H,2018ApJ...866...90C} and stretched the SMC along the line of sight, giving it an elongated shape \citep{2016AcA....66..149J, 2017MNRAS.472..808R}. Despite ongoing efforts, the dynamical history of the Magellanic system remains only partially understood. Various substructures observed in the system, originating from the gravitational interactions between the SMC, the LMC and the MW, offer key insights into its complex evolutionary past: 

\begin{itemize}
    \item {\itshape The Magellanic Stream} is a complex network of gas, filaments, and clumps extending over $200$ deg along the orbit of the MCs around the MW \citep{1974ApJ...190..291M, 2010ApJ...723.1618N}. This structure consists of material stripped from both the LMC and SMC \citep{2013AAS...22140404N, 2013ApJ...772..111R}. While some hydrodynamic simulations attribute its origin to ram-pressure from the MW halo \citep{2005MNRAS.363..509M, 2019MNRAS.486.5907W}, the prevailing view is that it formed from material stripped out from the MCs due to tidal interactions between them \citep{2016ARA&A..54..363D, 2018ApJ...857..101P, 2024Ap&SS.369..114L}.
    
    \item {\itshape The leading arm} stretches for $\sim50$ deg to the northeast of the LMC \citep{1998Natur.394..752P, 2003ApJ...586..170P}. It is thought to be the tidal counterpart of the Magellanic Stream, as its presence ahead of the MCs' orbit may only be explained by tidal forces.
    
    \item {\itshape The Magellanic Bridge} \citep{1963AuJPh..16..570H} is a structure that connects the LMC to the eastern extension of the SMC, called the 'Wing', and consists of both metal-poor gas and a stellar component \citep{1985Natur.318..160I}. Most studies attribute its formation to the most recent interaction between the two Clouds, with material primarily stripped from the SMC \citep{1996MNRAS.278..191G, 2012MNRAS.421.2109B, 2012ApJ...750...36D}. In addition to the H\,{\small I} gas distribution, the Bridge has been traced via its stellar content through its population of young blue stars \citep{1990AJ.....99..191I}, star clusters \citep[e.g.,][]{2020AJ....159...82B} and variable stars \citep{2020ApJ...889...26J,2020ApJ...889...25J}. Proper-motion studies reveal a net flow of stars from the SMC toward the LMC \citep{2019ApJ...874...78Z, 2020A&A...641A.134S, 2021A&A...649A...7G}. The Bridge hosts both young stars and candidate older populations, though the latter are difficult to confirm due to low spatial density and contamination from the MW \citep{2013A&A...551A..78B, 2013ApJ...779..145N, 2014ApJ...795..108S}. The most comprehensive study of the region’s SFH to date, by \citet{2007ApJ...658..345H}, found no evidence for an old population of tidally stripped stars in the Bridge. 

    \item {\itshape External substructures} 
    The MCs are surrounded by an extended network of low surface-brightness stellar features resulting from their dynamical history. Both N-body simulations and observations indicate the presence of tidally distorted stellar populations (e.g., the formation of spiral-like arms extending from the LMC disc \citep{2019MNRAS.482L...9B, 2022MNRAS.510..445C, 2022ApJ...931...19G}), likely induced by combined tidal influence of the SMC and the MW. In the SMC, the intermediate-age populations exhibit a pronounced elongation toward the Magellanic Bridge \citep{2019MNRAS.490.1076E}. 

\end{itemize}

Even if the orbital history of the MCs is only partially understood, our knowledge has improved significantly. High-precision Hubble Space Telescope (HST) measurements of the MCs' proper motions \citep{2006ApJ...638..772K, 2006ApJ...652.1213K} revealed that their velocities were too high for a system gravitationally bound to the MW. This overturned the traditional view of a long-standing bond (\( \gtrsim 10 \) Gyr) between the MW and the MCs \citep{1980PASJ...32..581M}. This led to the ‘first passage’ scenario \citep{2007ApJ...668..949B, 2012MNRAS.421.2109B, 2018MNRAS.473.1218L, 2021ApJ...921L..36L}, where the Magellanic system is on its first encounter with the MW\footnote{A recent study from \citet{2024MNRAS.527..437V} suggests that a two-passage scenario for the Magellanic Clouds remains possible, provided that the first passage occurred with a very large pericenter.}. This model is supported by the observed SF and high gas content in the MCs, which would have likely been stripped away in repeated interactions, hindering star formation (SF). Indeed, other HI-rich dwarf irregular satellites are typically found at greater distances from large spirals \citep{2003AJ....125.1926G, 2012AJ....144....4M}. The complex geometry of the MCs is more plausibly the result of interactions between the Clouds themselves \citep{2016ApJ...825...20B, 2022ApJ...927..153C}, supported by the presence of a common H\,{\small I} envelope \citep{2009IAUS..256...81V}. Models involving repeated SMC-LMC encounters \citep{2011AAS...21742404B, 2012MNRAS.421.2109B, 2014MNRAS.444.1759G, 2018ApJ...857..101P, 2019MNRAS.486.5907W} successfully reproduce their asymmetric features and gas structures, as well as  triggered bursts of SF observed in both galaxies. Their most recent close encounter is estimated to have occurred $150-250$ Myr ago \citep{2018ApJ...864...55Z, 2019ApJ...874...78Z, 2020A&A...641A.134S, 2020MNRAS.495...98D}, an event that N-body models \citep{1996MNRAS.278..191G, 2014MNRAS.444.1759G} predict led to the formation of the Bridge. 

Even though the number and timing of past interactions remain uncertain, analyzing their SFHs can reveal bursts of activity triggered by the gas compression from these encounters \citep{2015AAS...22521202S}. The SFH of the main bodies of the MCs has been extensively studied. Early investigations using the HST provided unprecedentedly deep photometry of spatially small, localized regions within both galaxies \citep[e.g.,][]{1999AJ....118.2262H, 2001ApJ...562..303D, 2009ApJ...703..721S, 2009AJ....137.3668C, 2012ApJ...754..130C, 2013ApJ...775...83C, 2013MNRAS.431..364W}. However, these detailed snapshots were limited by small-number statistics due to the small field of view of HST. The radial age gradient has different behaviors in the two galaxies. \citet{2014MNRAS.438.1067M} found that in the outer LMC disk younger populations are increasingly concentrated toward smaller radii, indicating an outside-in quenching pattern. This is consistent with \citet{2024ApJ...975...42C}, who show that the inner disk has an inside-out gradient that reverses beyond $\sim2$ scale-lengths, with older stars at larger radii, likely due to radial migration. In contrast, the SMC exhibits a classical outside-in gradient, with older stars in the outskirts and ongoing SF in the central Wing \citep{2024ApJ...975...43C}. The advent of wide-field surveys has since enabled the reconstruction of the global SFH of the MCs. Key surveys that have driven this effort include:

\begin{itemize}
    \item {\itshape Magellanic Clouds Photometric Survey (MCPS)} \citep{2002AJ....123..855Z, 2004AJ....128.1606Z} is the first large-scale study of stellar populations in the MCs. \citet{2004AJ....127.1531H} found that about half of the SMC's stars formed over $\sim8.4$ Gyr ago, followed by a prolonged period of quiescence. Subsequent bursts of SF occurred 2.5 Gyr, 400 Myr and 60 Myr ago, with counterparts at similar times in the LMC \citep{2009AJ....138.1243H}, indicating multiple past interactions;
    
    \item {\itshape VISTA survey of the Magellanic Clouds system (VMC)} \citep{2011A&A...527A.116C, 2025A&A...699A.300C} 
    is a near-IR survey; \citet{2018MNRAS.478.5017R} found that \(\sim 50\%\) of the stellar mass of the SMC formed over 6.3 Gyr ago, and confirmed that the older stellar population follows a smooth, elliptical distribution. The SMC's bar and Wing exhibit a complex SFH, with multiple recent bursts. While the western part of the SMC experienced a relatively constant SF up to $1.5$ Gyr ago, the eastern regions show two prominent peaks at 3.1 and 8 Gyr ago. The presence of stars younger than 200 Myr in the Wing supports the idea that a recent close encounter with the LMC triggered SF;
    
    \item {\itshape Survey of the Magellanic Stellar History (SMASH)} \citep{2017AJ....154..199N} provided further evidence of synchronized SF between the MCs. \citet{2022MNRAS.513L..40M} identified five peaks of SF in the SMC (3, 2, 1.1 and 0.45 Gyr ago, as well as an ongoing burst) aligned with bursts in the northern LMC \citep{2020A&A...639L...3R}, suggesting the galaxies synchronized their SF $\sim3.5$ Gyr ago, likely due to mutual gravitational interactions.
\end{itemize}

\begin{figure*}[h]
    \centering
    \includegraphics[width=0.98\textwidth,trim={10mm 0 30mm 28mm},clip]{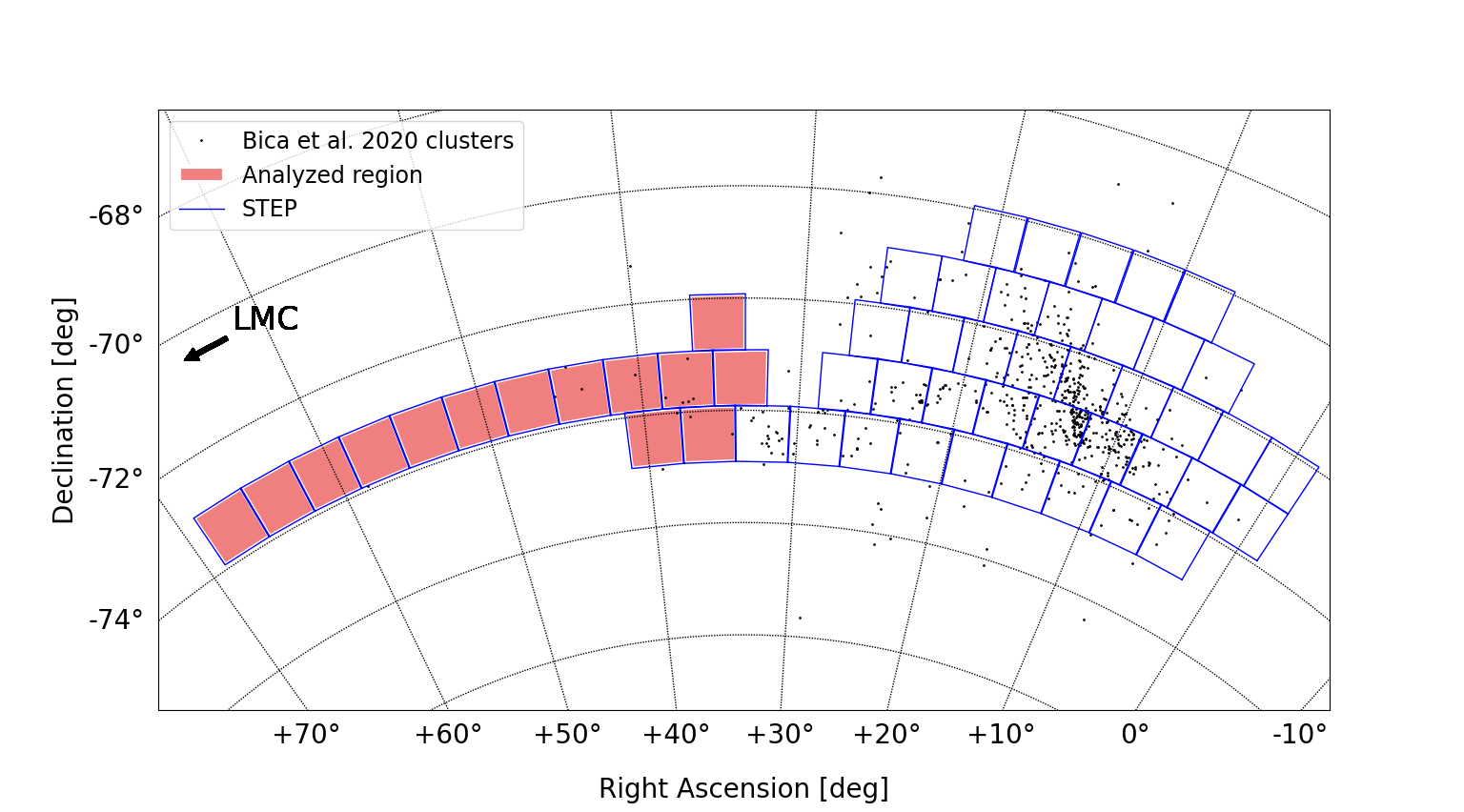}
    \caption{Footprint of the STEP survey pointings (blue boxes), showing the location of the SMC body and the Bridge. The area covered by this work is highlighted in red. The LMC is in the left direction, outside the map. The black dots mark the position of star clusters and associations from the catalogue of \citet{2020AJ....159...82B}.}
    \label{fig:mappa}
\end{figure*}

The goal of this paper is to derive the most extended SFH of the Magellanic Bridge employing deep optical data from the ``Small Magellanic Cloud in Time: Evolution of a Prototype interacting late-type dwarf galaxy'' \citep[STEP;][]{2014MNRAS.442.1897R} survey, obtained using the VLT Survey Telescope \citep[VST;][]{2011Msngr.146....2C}. The paper is structured as follows: Section 2 outlines the STEP survey data, its processing, and the artificial star tests conducted. Section 3 discusses the synthetic CMD method, detailing the construction of our synthetic models, including the distances and extinctions applied to all analyzed regions. In Section 4, we present the best-fit CMDs for selected regions and describe the results for the Magellanic Bridge as a whole. Finally, Section 5 summarizes the key findings and outlines the main conclusions.

\section{The Magellanic Clouds in STEP}

\subsection{STEP data}

STEP \citep[][]{2014MNRAS.442.1897R,Gatto2021} is a deep and extended photometric survey of the SMC and the Magellanic Bridge carried out with the 2.6m VLT Survey Telescope (VST), equipped with the wide-field imager Omegacam \citep{2011Msngr.146....2C, 2011Msngr.146....8K}. 
The STEP survey covered the main body of the SMC and the Magellanic Bridge using the g, r, i and H\(\alpha\) filters \citep{1996AJ....111.1748F, 1998AJ....116.3040G} at a limiting magnitude (\(g \sim 24 \) mag), which allows for the detection of stars well below the turn-off of the oldest population (\(g \sim 23\) mag in the SMC and \(\sim 22.5\) mag in the LMC) with a signal-to-noise ratio of 10 \citep{2014MNRAS.442.1897R}, even in the most crowded regions, an ideal condition to reliably recover the SFH at all epochs. The entire survey covers a total area of 53 $\mathrm{deg\, {^{2}}}$. In this work, we focused on the Magellanic Bridge. To this end, we selected 14 tiles\footnote{A tile is the result of the stacking of five Omegacam exposures properly dithered to cover the gaps between the 32 CCDs composing the camera. The typical sky coverage of each tile exceeds slightly 1 squared degree.}, highlighted as shaded regions in the map of Fig.~\ref{fig:mappa}, where the approximate position of the SMC body and the Bridge are marked by star clusters and associations from \citet{2020AJ....159...82B}.  The analyzed region of the Bridge extends from right ascension (\(\alpha\)) $4$h $28$m $59$s and declination (\(\delta\)) $-73$° $24$’ $59$” \citep[tile 3\_21, closer to the LMC, see][for the identification of the tiles]{2014MNRAS.442.1897R} to \(\alpha\) = $2$h $15$m $00$s and \(\delta\) = $-73$° $24$’ $59$” (tile 3\_11, on the SMC side).

\subsection{Data reduction and photometry}

The data reduction was performed using the VST-TUBE package \citep{2012MSAIS..19..362G}, specifically developed to process OmegaCAM data from the VST telescope. The pipeline includes pre-reduction, astrometric and photometric calibration, and mosaic production. Further details on the reduction process are provided in \citet{2014MNRAS.442.1897R}. The point spread function (PSF) photometry was carried out with the DAOPHOT IV/ALLSTAR package \citep{1987PASP...99..191S, 2011ascl.soft04011S}. To account for possible distortions at the edges of the tiles, the PSF function was allowed to vary across the field-of-view. The $\alpha$ and declination $\delta$ of each source were calculated using the package \textit{xy2sky}, a utility that converts the image coordinates $(X,Y)$ to equatorial coordinates using the world coordinate system (WCS) information in the image header. The absolute photometric calibration was obtained by using secondary standard stars in each tile as provided by the AAVSO Photometric All-Sky Survey (APASS) project\footnote{https://www.aavso.org/apass \citep{2015AAS...22533616H}} (see \citet{Gatto2020} for further details). The photometric catalog was cleaned of spurious objects, such as distant galaxies and artifacts caused by bad pixels, using the SHARPNESS parameter from the DAOPHOT package. This parameter measures how much broader a source appears compared to the PSF profile, allowing for the removal of extended sources from the catalog. The acceptance criteria for sources varied between tiles, with slightly different SHARP parameter ranges applied to each tile, usually around $\pm0.5$ (see Table \ref{tab:parameters} for the SHARPNESS cut values used in each tile.). We provide the final catalog, composed by $600279$ stars located within the Magellanic Bridge, for the scientific community. Table \ref{tab:catalog} offers an overview of the star-catalog, whose full version is available at the CDS. To isolate the field stellar population for a cleaner analysis, we excluded stars associated with known clusters by cross-matching our data with the \citet{2020AJ....159...82B} catalog. A total of 13 clusters were removed, with stars excised if they fell within the major angular radii listed in the catalog (ranging from $21''$ to $1.2'$, except for one cluster in tile 3\_15, which extended to $2'$ $27''$). This process eliminated 225 stars from our data, averaging $\sim20$ stars per cluster, with a minimum of 5 and a maximum of 42 stars.

\begin{table*}[htbp] 
\centering
\footnotesize
\caption{Photometric catalog of the STEP survey in the Magellanic Bridge}
\label{tab:catalog}

\begin{tabular*}{0.98\textwidth}{@{\extracolsep{\fill}}ccccccccc@{\extracolsep{\fill}}}

\hline\hline
{ID} & {RA} & {Dec} & {$g$} & {$\sigma_g$} & {$i$} & {$\sigma_i$} & {$CHI$} & {$SHARP$} \\
{} & {(deg)} & {(deg)} & {(deg)} & {(mag)} & {(mag)} & {(mag)} & {} & {} \\
\hline
STEP-J022137.74-725610.7 & 35.40724 & -72.93630 & 22.106 & 0.005 & 20.933 & 0.006 & 0.80 & -0.19 \\
STEP-J022140.41-725637.4 & 35.41838 & -72.94373 & 22.696 & 0.009 & 22.022 & 0.014 & 0.76 & -0.11 \\
STEP-J022140.24-725425.4 & 35.41768 & -72.90705 & 20.912 & 0.002 & 18.221 & 0.001 & 0.96 & -0.05 \\
STEP-J022140.02-725753.2 & 35.41676 & -72.96477 & 21.230 & 0.003 & 18.870 & 0.001 & 1.07 & -0.19 \\
STEP-J022132.34-725505.7 & 35.38476 & -72.91826 & 22.719 & 0.010 & 21.055 & 0.007 & 0.89 & -0.11 \\
STEP-J022147.68-725458.6 & 35.44864 & -72.91628 & 21.050 & 0.003 & 19.896 & 0.003 & 1.03 & 0.20 \\
STEP-J022136.05-725423.2 & 35.40022 & -72.90644 & 22.689 & 0.009 & 22.159 & 0.019 & 0.75 & -0.11 \\
STEP-J022148.05-725407.1 & 35.45019 & -72.90197 & 21.888 & 0.005 & 21.855 & 0.014 & 0.76 & 0.12 \\
STEP-J022144.25-725418.2 & 35.43437 & -72.90505 & 23.373 & 0.015 & 21.625 & 0.012 & 0.71 & 0.05 \\
\hline
\end{tabular*}

\tablefoot{
Columns contain: (1) object identifier; (2)--(3) celestial coordinates for epoch J2000; (4)--(5) PSF-calibrated $g$-band magnitudes and uncertainties; (6)--(7) $i$-band counterparts; (8) DAOPHOT CHI parameter; (9) DAOPHOT SHARP parameter. A portion is shown here for guidance regarding its form and content. The machine-readable version of the full table will be published at the Centre de Données astronomiques de Strasbourg (CDS, \url{https://cds-u-strasbg.fr/}).
}

\end{table*}

\subsection{Artificial star tests}
\label{sec:AST}

\begin{figure}[]
    \centering
    \includegraphics[width=0.5\textwidth]{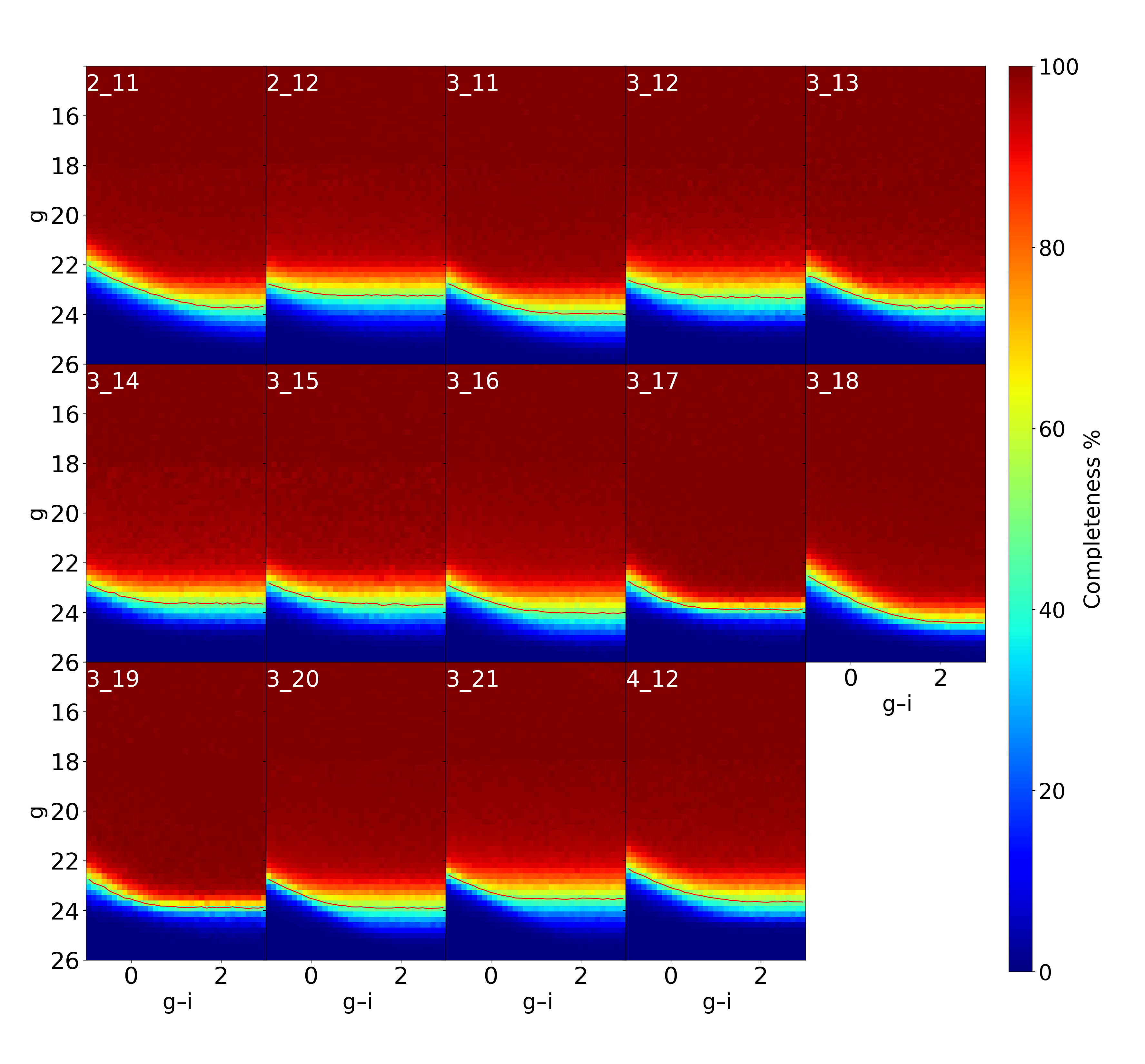}
    \caption{Hess diagrams of the completeness for all analyzed tiles. In each cell of the CMD the completeness is computed as the ratio of recovered artificial stars with respect to the number of input stars. The red line indicates the 50\% completeness level.}
    \label{fig:compl}
\end{figure}

\begin{figure}[]
    \centering
    \includegraphics[width=0.45\textwidth,trim={30mm 5mm 30mm 30mm},clip]{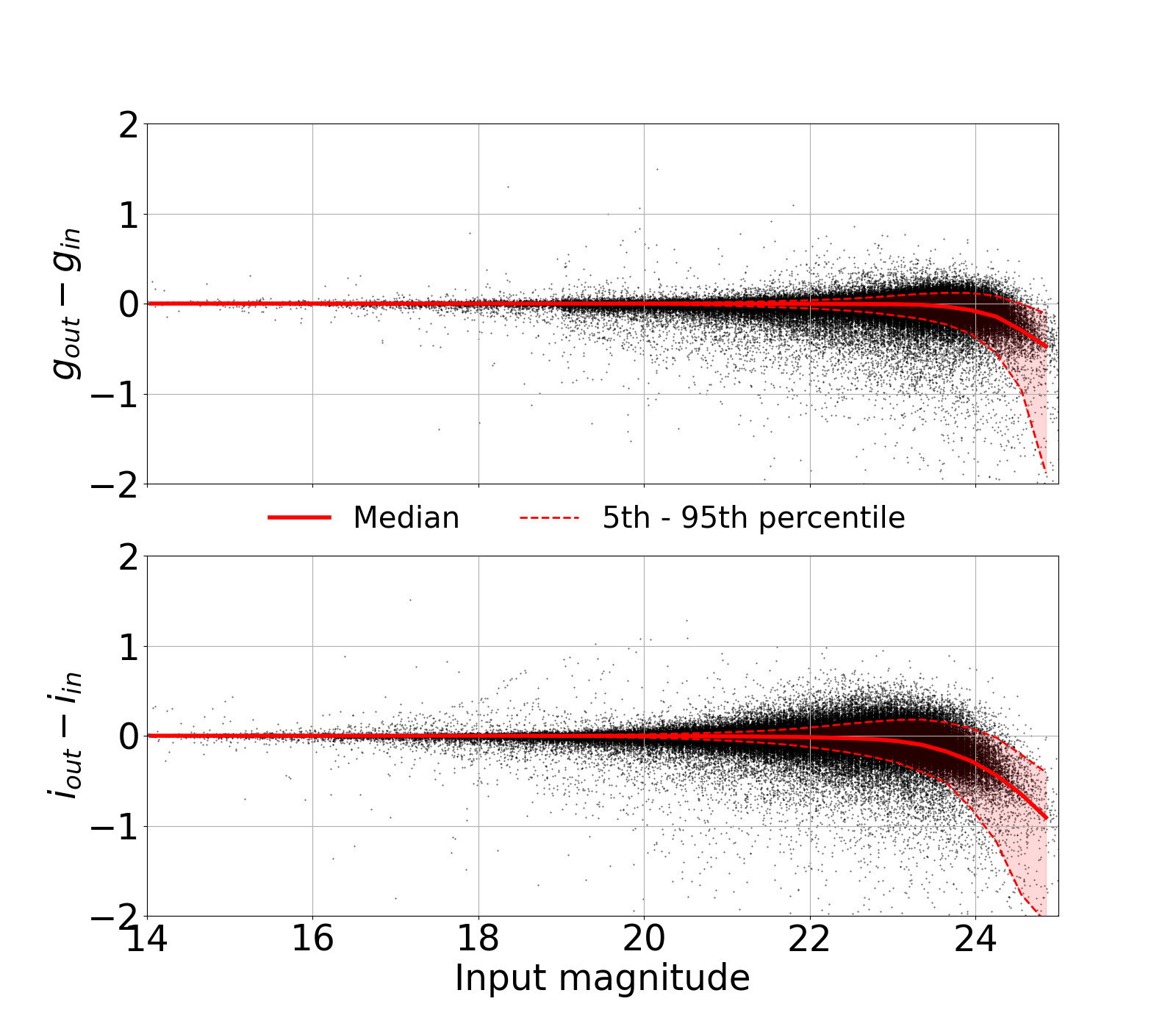}
    \caption{Difference between the output and input magnitude vs. input magnitude in $g$ (top) and $i$ (bottom) filters derived from AS tests for tile $3\_17$. The red solid line marks the median of the distributions, while the dashed lines indicate the \(5^{th}\) and \(95^{th}\) percentiles of the distributions.}
    \label{fig:error}
\end{figure}

Two key factors for the reliability of the SFH estimated from PSF photometry are: the photometric error and the incompleteness of the stellar photometry \citep{2010AdAst2010E...3C}. The only accurate way to estimate these quantities across our magnitude range is by repeating the entire process of photometric measures on the science images on which artificial stars (AS), of known position and magnitudes, are injected. To achieve a statistically reliable result, we generated a large number of artificial stars ($\sim 2-3\times10^6$) with the appropriate Poisson noise and PSF, injecting them into each tile. These stars fully sample the range of magnitudes and colors covered by the data, spanning from $g=14$ mag to $26$ mag and a $g-i$ color range between $-1.0$ mag and $3.0$ mag. The magnitude distribution of artificial stars is not uniform: 80\% of them have $g>19$ mag to better map completeness and photometric errors in the fainter regions of the CMD. Moreover, to avoid introducing additional crowding, the artificial stars are placed in the image on a grid with mutual distances much larger than the PSF, typically $10''$ in our tests. The images containing real and artificial stars are then re-analyzed with the DAOPHOT package. 
An artificial star is considered successfully recovered if its estimated position differs by less than 1 pixel ($\sim 0.2''$) from the input value, and if the recovered magnitude is not more than 0.75 mag brighter than the input one. In addition, we applied the same criteria for the SHARPNESS parameter to the artificial stars as we did for the real data. The completeness of our photometry was computed as the percentage of recovered artificial stars within each magnitude and color interval, while photometric errors were derived as the difference in magnitude between the recovered and input artificial star in each filter. Our tests show that, across all regions of the Bridge, the STEP survey reaches 50\% completeness at $g \sim 23.5$ mag and $i \sim 23$ mag, and at slightly fainter values in less crowded regions. To ensure a completeness level $\gtrsim 50\%$, we applied a magnitude cut at $g = 23.2$ mag in all tiles in the recovery of the SFH. The estimated photometric error at the magnitude of the turn-off of the oldest population in the SMC ($\sim 22.8$) is $\sim 0.2$ mag in the $g$ filter. The fraction of recovered stars is best visualized in the CMD. By binning the CMD, we can construct two-dimensional histograms (‘Hess diagrams’) that depict the frequency of stars in each color–magnitude cell. Figure \ref{fig:compl} shows the Hess diagrams with the completeness distributions of all tiles, where the 50\% completeness level (typically at $g\sim 23.2$ mag for $g-i=0$) is marked by a red line. To obtain a smooth distribution, we used a relatively large cell size of 0.1 mag in color and 0.2 mag in magnitude. Figure \ref{fig:error} illustrates the photometric error as a function of the input magnitude in $g$ and $i$, where the central red line marks the median of the distribution, which becomes negative for fainter stars due to the effects of blending.

\section{The SFH recovery method}

A complex mix of heterogeneous stellar populations can be analyzed comparing the data against synthetic populations, employing the well-established ‘synthetic CMD’ approach (see e.g., \citet{1991AJ....102..951T, 1996ApJ...462..672T, 1997NewA....2..397D, 2002MNRAS.332...91D, 1999MNRAS.304..705H, 2009AJ....138..558A, 2010AdAst2010E...3C, 2012ApJ...748...88W}). The SFH recovery was performed employing SFERA \citep[Star Formation Evolutionary Recovery Algorithm][]{2009AJ....137.3668C, 2015ApJ...811...76C}, which utilizes Pikaia \citep{1995ApJS..101..309C}, a genetic algorithm that efficiently explores the parameter space independently of initial conditions.
In our work, we built two catalogs of isochrones covering the age range from $10$ Myr to $13.4$ Gyr and [Fe/H] from -2.0 to 0.0 dex. For our study, we adopted the PARSEC-COLIBRI \citep{2012MNRAS.427..127B, 2017ApJ...835...77M} and BaSTI \citep{2018ApJ...856..125H} stellar isochrones. In both sets of models we used solar-scaled abundances, convective overshooting, and a Reimers mass-loss parameter of $\eta = 0.3$. We grouped these isochrones into quasi-simple stellar populations (QSSPs or partial models), representing finite-duration episodes of constant SF. Each QSSP covers a specific time-span (see Table~\ref{tab:eta_durata}) and includes stars uniformly distributed within a small metallicity interval (0.1 dex). 
To populate the QSSPs, we employed a Monte Carlo approach, randomly selecting metallicity, age, and mass. The effective temperature \(T_{\rm eff}\) and luminosity L of each synthetic star were then assigned by interpolating along the isochrone of appropriate age. To avoid under-populating the upper end of the mass spectrum, mass values were drawn from a uniform distribution between the hydrogen-burning limit ($0.1 {M_\odot}$) and the maximum mass that is still alive at the corresponding model age. 
Each star was assigned a statistical weight, based on its formation probability derived from a \citet{2001MNRAS.322..231K} initial mass function (IMF). The presence of unresolved binaries alters the morphology of the CMD and we need to include this effect in our models: 30\% of the synthetic stars were assumed to be unresolved binaries and the mass of the secondary star was randomly extracted from a \citet{2001MNRAS.322..231K} IMF. In the creation of binaries, we only simulate the photometric effect of seeing the fluxes of two stars as one single object when we populate the synthetic model. To ensure a proper representation of all stellar evolutionary phases, including rapid ones, each model was populated with a large number ($10^6$) of synthetic stars. In the end, we generated two libraries, each containing partial models of 19 different ages and 20 different metallicities (380 per library). Before comparing these models with our observations, we convolved them with the data characteristics, as described in the following sections.

\begin{table}
\centering
\caption{Age bins used for the SFH analysis and their durations.}
\footnotesize\setlength{\tabcolsep}{3.0pt}
\label{tab:eta_durata}
\begin{tabular}{cc|cc|cc}
\hline\hline
{Final Age} & {Duration} & {Final Age} & {Duration} & {Final age} & {Duration} \\
\hline
25 Myr & 15 Myr & 400 Myr & 100 Myr & 3.50 Gyr & 0.75 Gyr \\
50 Myr & 25 Myr & 500 Myr & 100 Myr & 5.00 Gyr & 1.50 Gyr \\
75 Myr & 25 Myr & 750 Myr & 250 Myr & 7.00 Gyr & 2.00 Gyr \\
100 Myr & 25 Myr & 1.00 Gyr & 250 Myr & 9.00 Gyr & 2.00 Gyr \\
150 Myr & 50 Myr & 1.50 Gyr & 0.50 Gyr & 13.40 Gyr & 4.40 Gyr \\
200 Myr & 50 Myr & 2.00 Gyr & 0.50 Gyr \\
300 Myr & 100 Myr & 2.75 Gyr & 0.75 Gyr \\
\hline
\end{tabular}
\end{table}

\subsection{Distance and extinction}

We adapted the libraries of partial models to match the characteristics of each observed tile. For the distance, we initially assumed a linear gradient between the distance moduli of the LMC and SMC found in the literature, $\sim18.47$ mag \citep{2019Natur.567..200P} and $\sim 18.97$ mag \citep{2020ApJ...904...13G}, respectively. For the Galactic foreground reddening correction, we adopted E(B-V) values from the SFD reddening map \citep{2011ApJ...737..103S}. We employed the \citet{1989ApJ...345..245C} extinction law, assuming a normal total-to-selective extinction value of $R_V=3.1$. To account for differential reddening within each region, we determined a unique reddening value for every synthetic star that was randomly selected from a Gaussian distribution, centered on the median E(B-V) and its spread for that particular region. Additionally, we applied a stronger reddening correction for younger stars ($\le75$ Myr). For these stars, the E(B-V) value was twice that applied to older stars. The final parameters chosen in the analysis of each tile are listed in Table \ref{tab:parameters}. 
The extinction toward the Bridge is generally low, with a median E(B-V) value of $\sim0.06$ mag, in line with previous works on this low dust region \citep{2017MNRAS.466.4138W}. The extinction increases slightly toward the dustier outskirts of the LMC.

\begin{table*}[]
\centering
\caption{Summary of the parameters used in the recovery of the SFH of each tile in the Magellanic Bridge.}
\small 
\setlength{\tabcolsep}{6pt} 
\begin{tabular}{lccrcccccccc}
\hline\hline
{Tile} & \multicolumn{2}{c}{Coordinates} & {\makecell{Number\\of stars}} & \multicolumn{2}{c}{E(B-V)} & {\makecell{Distance\\ Modulus}} & \multicolumn{2}{c}{FWHM} & {\makecell{Phot. error at \\g=22.8 mag}} & \multicolumn{2}{c}{{SHARPNESS}} \\
\cmidrule(lr){2-3} \cmidrule(lr){5-6} \cmidrule(lr){8-9} \cmidrule(lr){11-12}
 & {RA [h:m:s]} & {Dec} & & {mean} & {dispersion} & (mag) & {g ('')} & {i ('')} & (mag) & {min} & {max} \\
\hline
2\_11 & 02:23:27 & -74° 24' 29" & 16,227 & 0.050 & 0.005 & 18.82 & 0.99,1.20 & 0.59,1.03 & 0.03 & -0.4 & 0.5 \\
2\_12 & 02:37:39 & -74° 24' 29" & 14,615 & 0.060 & 0.007 & 18.85 & 1.06,1.47 & 0.95,1.08 & 0.03 & -0.5 & 0.5 \\
3\_11 & 02:15:00 & -73° 24' 59" & 20,419 & 0.060 & 0.008 & 18.86 & 0.98,1.13 & 1.03,1.07 & 0.03 & -0.4 & 0.5 \\
3\_12 & 02:28:24 & -73° 24' 59" & 11,398 & 0.040 & 0.007 & 18.84 & 1.07,1.17 & 1.02,0.99 & 0.04 & -0.3 & 0.4 \\
3\_13 & 02:41:47 & -73° 24' 59" & 11,510 & 0.060 & 0.008 & 18.77 & 1.22,1.23 & 1.06,1.04 & 0.05 & -1.0 & 1.0 \\
3\_14 & 02:55:11 & -73° 24' 59" & 14,061 & 0.040 & 0.008 & 18.75 & 1.12,1.17 & 0.95,1.02 & 0.04 & -0.4 & 0.5 \\
3\_15 & 03:08:35 & -73° 24' 59" & 14,274 & 0.035 & 0.008 & 18.73 & 1.23,1.25 & 0.99,0.96 & 0.03 & -0.4 & 0.55 \\
3\_16 & 03:21:58 & -73° 24' 59" & 17,664 & 0.055 & 0.010 & 18.66 & 1.29,1.74 & 0.83,0.96 & 0.02 & -0.4 & 0.5 \\
3\_17 & 03:35:22 & -73° 24' 59" & 18,739 & 0.065 & 0.010 & 18.58 & 1.19,1.15 & 0.95,1.02 & 0.04 & -1.0 & 1.0 \\
3\_18 & 03:48:46 & -73° 24' 59" & 21,277 & 0.075 & 0.020 & 18.55 & 1.23,1.01 & 1.16,1.12 & 0.03 & -1.0 & 1.0 \\
3\_19 & 04:02:10 & -73° 24' 59" & 27,946 & 0.090 & 0.020 & 18.50 & 1.13,1.22 & 0.96,0.96 & 0.04 & -1.0 & 1.0 \\
3\_20 & 04:15:34 & -73° 24' 59" & 68,022 & 0.104 & 0.020 & 18.45 & 1.37,1.19 & 0.75,0.73 & 0.03 & -0.5 & 0.6 \\
3\_21 & 04:28:58 & -73° 24' 59" & 155,884 & 0.120 & 0.030 & 18.45 & 1.36,1.41 & 0.63,0.78 & 0.05 & -0.55 & 0.65 \\
4\_12 & 02:20:10 & -72° 25' 27" & 11,670 & 0.045 & 0.008 & 18.86 & 1.05,1.21 & 0.91,1.00 & 0.03 & -0.45 & 0.45 \\

\hline
\end{tabular}
\tablefoot{Columns provide: (1) tile name; (2)--(3) J2000 equatorial coordinates; (4) total number of stars; (5)--(6) mean E(B-V) extinction and its dispersion; (7) distance modulus (m-M)$_0$; (8)--(9) FWHM in the g and i filters; (10) photometric error at g = $22.8$ mag; (11)--(12) sharpness range for source selection.}
\label{tab:parameters}
\end{table*}

\subsection{Photometric errors and incompleteness}

Our partial models were then convolved with the incompleteness and photometric errors derived from the AS tests described in Section \ref{sec:AST}. The completeness limit defines the faintest magnitude at which we can confidently detect and measure objects in our observed field. Ideally, we aim for a limit that allows us to observe main-sequence (MS) stars down to $0.8 {M_\odot}$, as these stars have lifetimes longer than the age of the universe tracing SF across cosmic time. 
The completeness distribution in the CMD determines the weight assigned to each synthetic star. Meanwhile, photometric errors, estimated as the difference between the recovered and input magnitudes, are used to model the uncertainty across the full magnitude range. An appropriate photometric error is assigned to each synthetic star based on its position in the CMD. To this purpose, we applied a bilinear interpolation scheme that displaces each synthetic star in both colour and magnitude, considering the error distribution from adjacent CMD cells. This approach ensures a realistic representation of observational uncertainties in our final models.

\subsection{Milky Way contamination}

\begin{figure}[]
    \centering
    \includegraphics[width=0.35\textwidth]{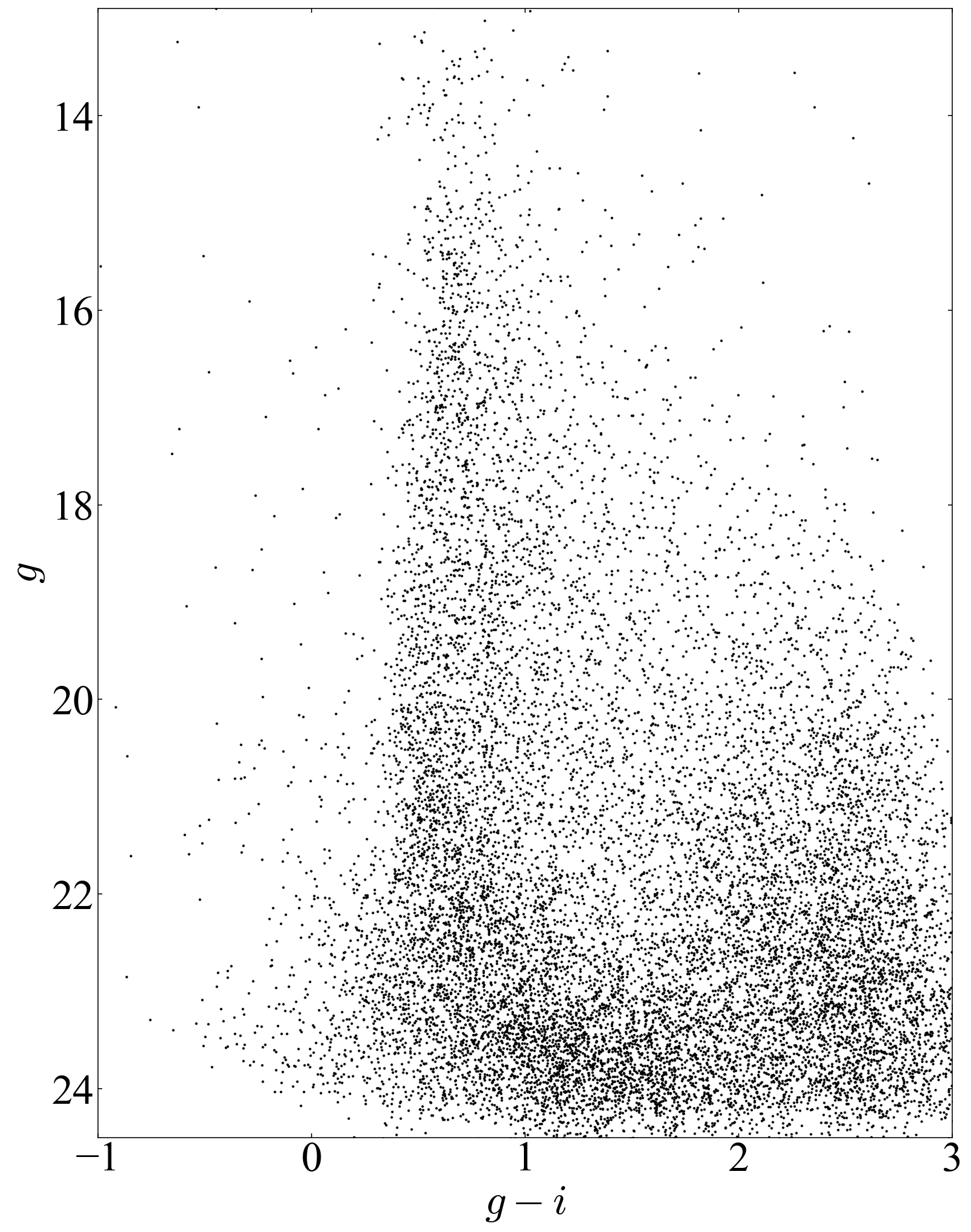}
    \caption{CMD of the tile 4\_14 from the YMCA survey, used to model the MW contamination in our analysis.}
    \label{fig:MW_CMD}
\end{figure}

To correct for MW contamination in our CMDs, we used a secondary field located near the Bridge (Figure \ref{fig:MW_CMD}). This field\footnote{Tile 4\_14, centered at RA = 02h 43m 54.6s and Dec = -70° 40' 41.52''} was observed as part of the Yes, Magellanic Clouds Again survey \citep[YMCA;][]{2024A&A...690A.164G}, a survey of the outskirts of the MCs from the same instrument, same filters, and with similar observing strategy. The stellar density of the LMC drops abruptly beyond approximately $9$° from its center and from $\sim6$° from the SMC center \citep{2024A&A...690A.164G}. We selected tile 4\_14 due to its relatively large distance from both the LMC ($\sim 13.4$°) and the SMC ($\sim 9.2$°), where the CMD is expected to be dominated by MW foreground stars (third panel, third row of Figure 9 in \citet{2024A&A...690A.164G}). This YMCA tile also shares a similar Galactic latitude with the analyzed regions. We excluded the presence of gradients in the color distribution of MW contaminants across our Galactic latitude range using TRILEGAL \citep{2005A&A...436..895G, 2012ASSP...26..165G} to simulate MW stellar counts across the range $-44$° $\leq b \leq -38$°, resulting in no sensible variation. An alternative approach would have been to generate the synthetic foreground CMD with TRILEGAL  for fields centered on the analyzed tiles. We also plot the Gaia DR3 data distribution for a region centered at $l=298°$ deg and $b=-44°$ deg, for reference. In appendix \ref{sec:MW_contamination} we present the comparison between the recovered SFH and the residuals obtained using the two approaches. We found that the synthetic Milky Way model generated with TRILEGAL produces significantly higher residuals, making it less effective in modeling contamination compared to secondary field observations. In light of these tests, we chose to use the YMCA tile as a representative case of MW contamination in all analyzed CMDs. In the end, The additional MW model is added to our partial model library and we incorporate it into the final synthetic population as described in the following paragraph.

\begin{figure}[]
    \centering
    \includegraphics[width=0.47\textwidth]{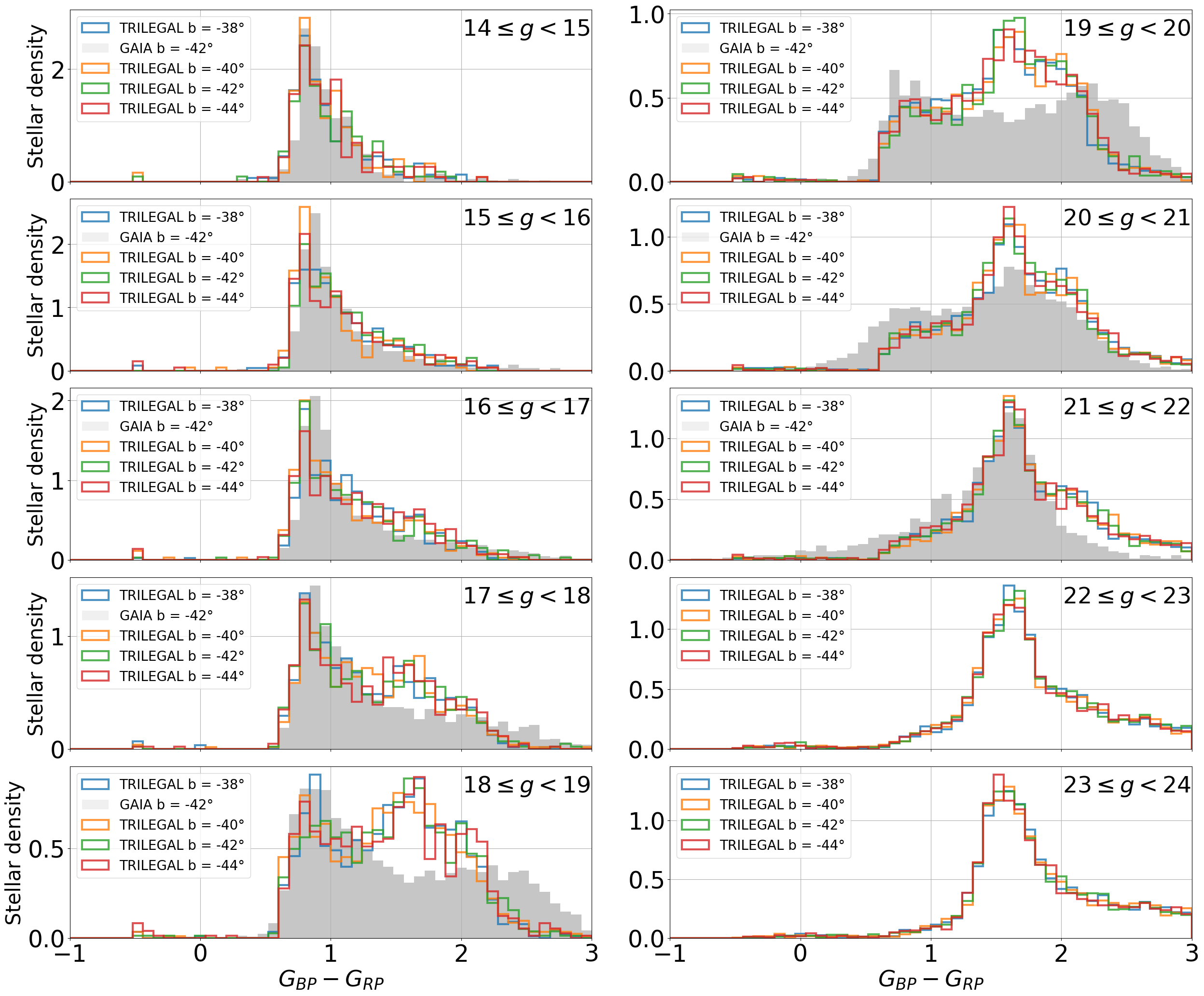}
    \caption{Stellar density distribution for Galactic longitude $l=289^\circ$ across different latitude bins ($-44$° $\leq b \leq -38$°). Each panel displays the normalized $G_{BP}-G_{RP}$ color distribution for different magnitude intervals sourced from the TRILEGAL Galactic model \citep{2005A&A...436..895G, 2012ASSP...26..165G}. The Gaia EDR3 distributions at $b=-42°$ are plotted in gray.}
    \label{fig:MW_dist}
\end{figure}

\subsection{Finding the best-fit model}
\label{sec:bestfit}

The next step is to construct a synthetic representation of a stellar population with a given SFH and age-metallicity relation (AMR) with a combination of the partial models. To quantify the similarity between a synthetic model and the observational data, we divide the CMDs into a grid of magnitude and color bins, then, for each bin, we compare the predicted and observed star counts. Once binned, the CMDs become Hess diagrams, where the third axis represents the number of stars inside each color-magnitude cell i. The Hess diagram of any synthetic model is therefore a linear combination of the $N\times M$ Hess diagrams of the partial CMDs:

\begin{equation}
MOD_i = \sum_{j=1}^{N} \sum_{k=1}^{M} w(j,k) \cdot p_i(j,k)
\label{eq:mod}
\end{equation}
Where \(MOD_i\) is the number of stars in cell i in the synthetic model, \(p_i(j,k)\) is the number of stars into cell i associated with the partial model of age step j and metallicity step k, and \(w(j,k)\) are the weighting coefficients associated to each partial model in the linear combination. The foreground model is added to the library of partial CMDs to estimate the contribution of sources of the MW. Thus, when we create the synthetic population, equation \eqref{eq:mod} becomes

 \begin{equation}
MOD_i = \sum_{j=1}^{N} \sum_{k=1}^{M} w(j,k) \cdot p_i(j,k) + w_{MW}\cdot f_i
\label{eq:mod2}
\end{equation}
Where \(w_{MW}\) is the weight associated with the MW model and \(f_i\) are the star-counts in bin i of the foreground CMD.
The statistical similarity between two Hess diagrams is evaluated with a Poisson based likelihood function, of the form

\begin{equation}
\chi^2_P = \sum_{i=1}^{N_{\text{cells}}} \left( \text{obs}_i \cdot \ln \frac{\text{obs}_i}{MOD_i} - \text{obs}_i + MOD_i \right)
\label{eq:chi}
\end{equation}
Where \(obs_i\) represent the number of star-counts in cell i in the data. The most likely SFH is determined by finding the combination of partial models that minimize the likelihood function of the residuals. To simplify the parameter space, in each tile, we consider populations with one value of distance modulus and mean extinction. We assess the sensitivity of the fit to these parameters by systematically varying them and repeating the SFH recovery. To optimize the comparison between models and observations, we employ a variable cell width in the CMD grid: finer bins where the CMD is densely populated; coarser bins where the stellar density is lower or models are more uncertain. The SFERA algorithm determines the optimal set of weighting parameters \(w(j,k)\) that generate the best-fitting synthetic stellar population. For each age step $t_j$ and metallicity step $Z_k$, the corresponding partial model of mass $M(j,k)$ has an associated weight $w(j,k)$. The total stellar mass formed in the age interval $t_j$ is then obtained by summing over all metallicity bins:

\begin{equation}
M_{\rm tot}(t_j) = \sum_{k=1}^{N} w(j,k)\, M(j,k),
\end{equation}
where $N$ is the number of metallicity steps. Dividing this mass by the duration of the age interval $\Delta t_j$ gives the SFR:

\begin{equation}
\mathrm{SFR}(t_j) = \frac{M_{\rm tot}(t_j)}{\Delta t_j} = \frac{\sum_{k=1}^{N} w(j,k)\, M(j,k)}{\Delta t_j}.
\end{equation}

Concerning the AMR, the mean metallicity each age step \(t_j\) is computed by summing the metallicity values of partial models, weighted by the ratio of their masses to the total mass of the best-fit model at that age.

\begin{equation}
Z(t_j) = \sum_{k=1}^{M} Z_k \cdot w(j,k) \frac{ M(j,k)}{\sum_{k=1}^{M}  w(j,k) M(j,k)}
\label{eq:Z}
\end{equation}
We repeat this procedure over all ages covered by the partial models to estimate the AMR. 

To assess the uncertainty around the best-fit model, we employ a bootstrapping technique, using 25 bootstrap resamplings per analyzed tile. After recovering the SFH from each bootstrap, we calculate the confidence interval around the solution as the \(5^{th}\) and \(95^{th}\) percentiles of the parameters’ distribution. The goodness of fit can be visually evaluated as the difference between the Hess diagrams of the observed data and the best-fit model. Ideally, the residuals should not exhibit systematic patterns but rather appear as structureless noise.

\section{Results}

\begin{figure}[]
    \centering
    \begin{subfigure}[b]{0.5\textwidth}
        \includegraphics[width=\textwidth]{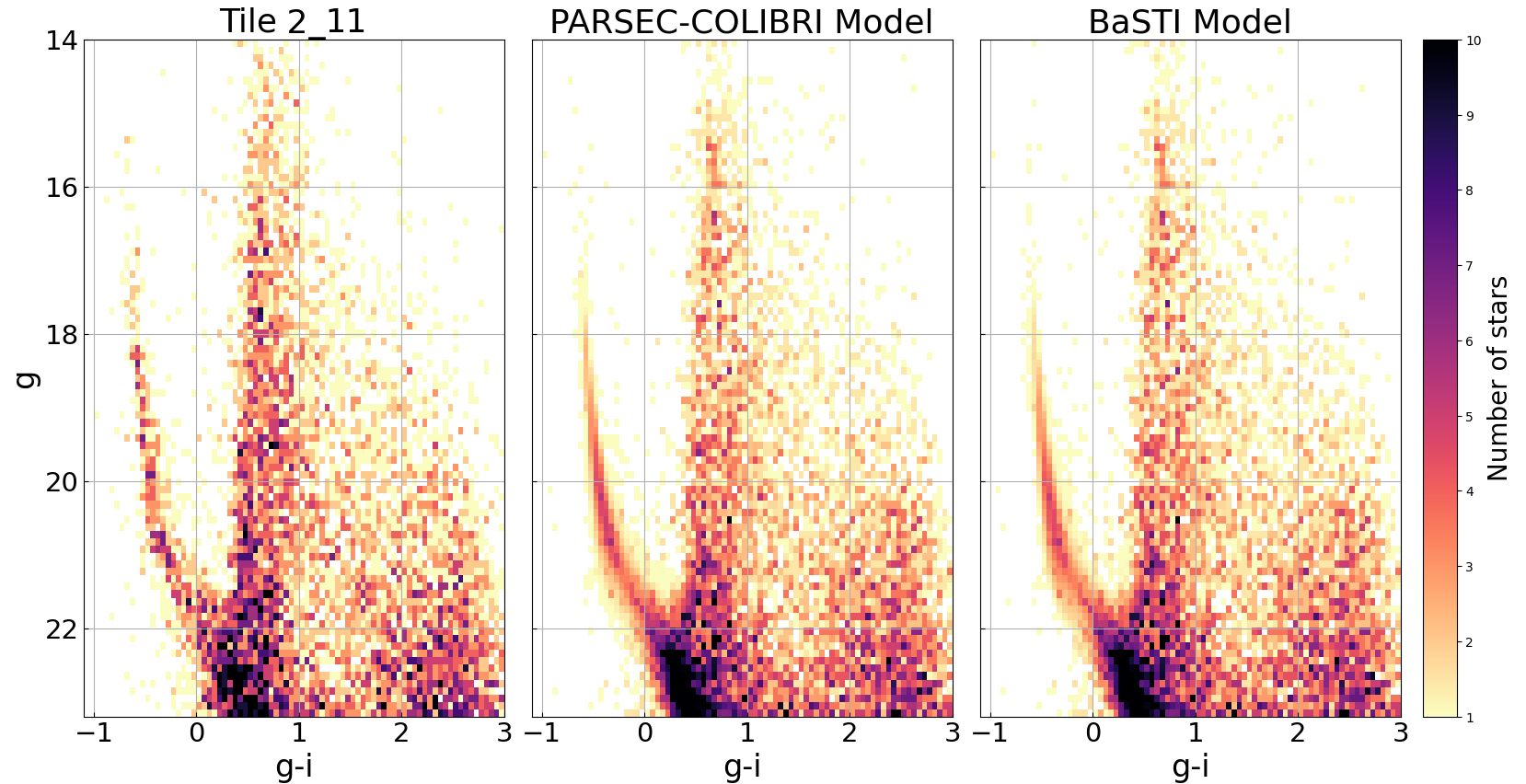}

        \label{fig:immagine1}
    \end{subfigure}
    \hfill
    \begin{subfigure}[b]{0.46\textwidth}
        \includegraphics[width=\textwidth, trim={8mm 0mm 0mm 0mm},clip]{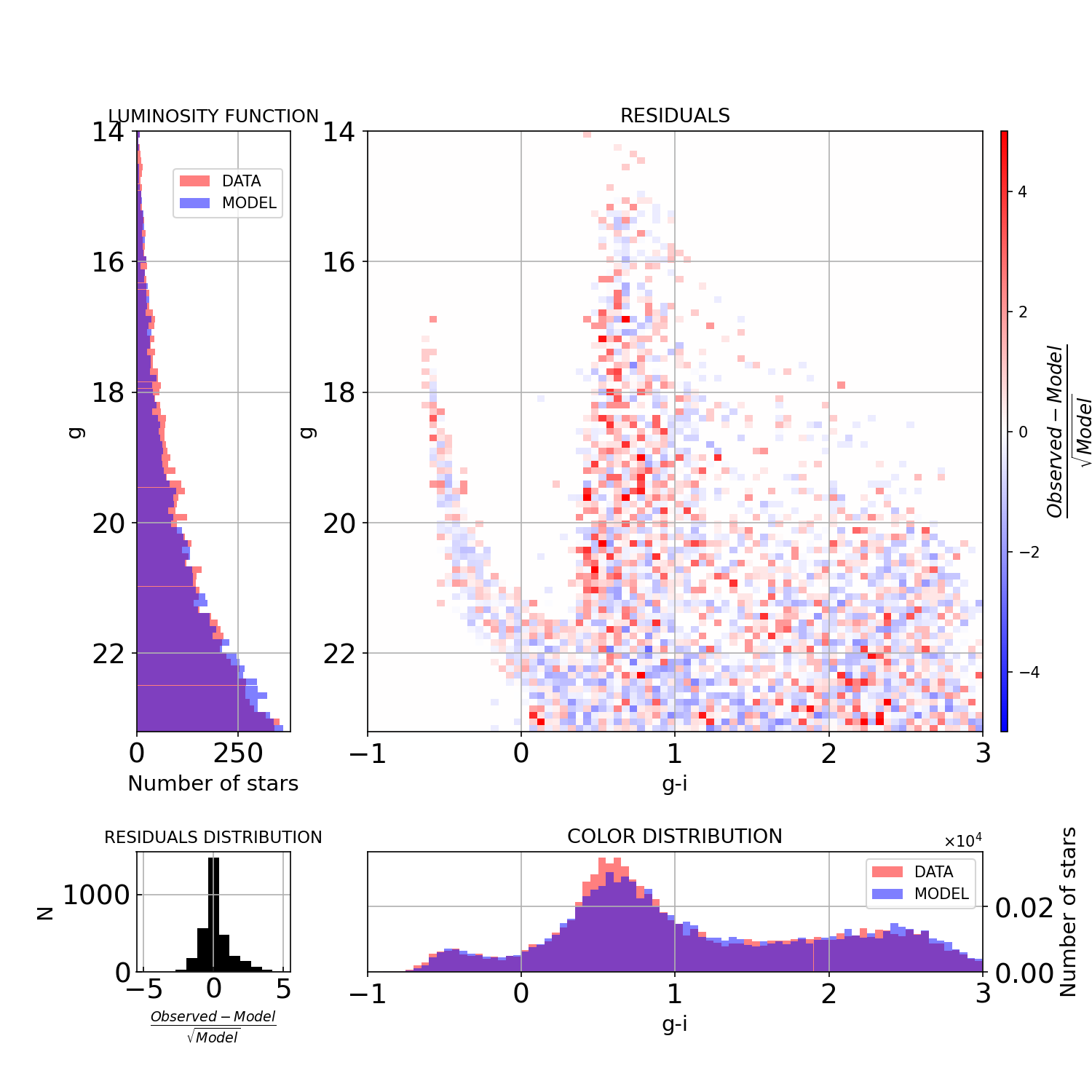}

        \label{fig:immagine2}
    \end{subfigure}
    \caption{CMDs of the observed data (top left) of tile 2\_11 compared with the best-fit synthetic model from PARSEC-COLIBRI (top center) and BaSTI (top right) stellar isochrones. In the plot at the bottom we show the residuals and the comparison between the observed and recovered luminosity functions and color distributions, as well as the distribution of residuals from the PARSEC-COLIBRI models.}
    \label{fig:models}
\end{figure}

\begin{figure}[h]
    \centering
    \includegraphics[width=0.45\textwidth, trim={0 0 20mm 10mm},clip]{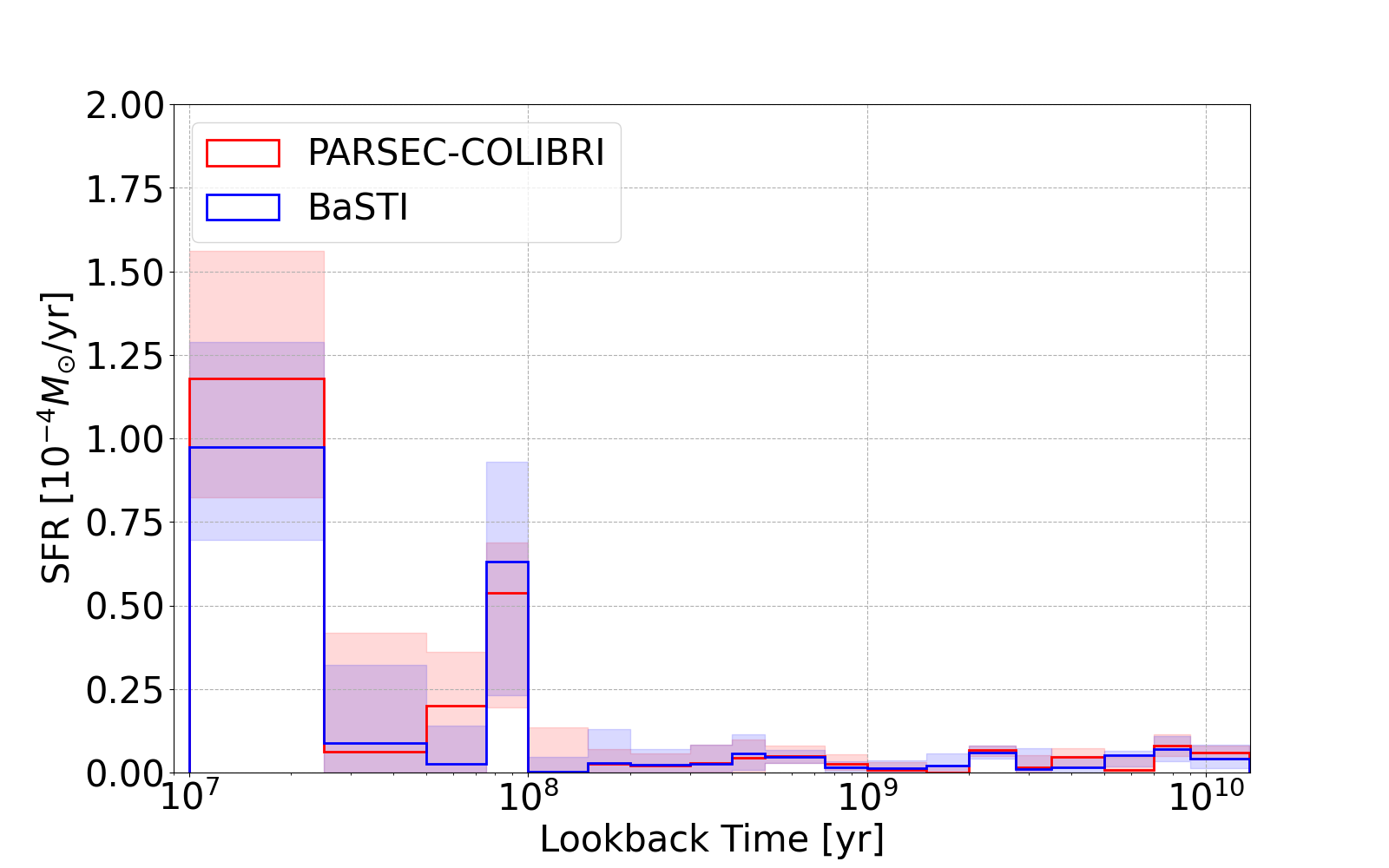}
    \caption{SFH of tile $2\_11$ recovered with PARSEC-COLIBRI (red) and BaSTI (blue) stellar isochrones.}
    \label{fig:211SFR}
\end{figure}

We present here the best-fit SFH for all tiles in the Bridge. As an example, Fig.~\ref{fig:models} (top panels) displays the observational data for tile 2\_11 alongside the best-fit synthetic models from PARSEC-COLIBRI and BaSTI. In the bottom panel, the figure also presents the residuals, the observed and recovered luminosity functions and color distributions.
The algorithm effectively captures the overall structure of the observed Hess diagram, successfully replicating all the characteristics of the observed CMDs, albeit with some remaining discrepancies in some regions, like the incorrect recovery of the red clump (RC) morphology in tile 3\_20 (see Figure \ref{fig:confronto320}). These residuals may be due to the theoretical modeling uncertainties of advanced stages of stellar evolution, particularly those involving the intricate interplay of physical processes that are not yet fully understood (such as late-phase stellar winds, convection, and mass loss). Nevertheless, ~90\% of the residuals fall within $\pm 1$ $\sigma$. In Figure \ref{fig:211SFR} we compare the SFHs obtained using the PARSEC-COLIBRI and BaSTI models in tile 2\_11. The results from the two stellar evolutionary libraries are, overall, in very good agreement, within the error bars, and this is also seen in almost all other tiles. 

\begin{figure*}[h]
    \centering
    \includegraphics[width=0.95\textwidth,trim={45mm 10mm 20mm 10mm},clip]{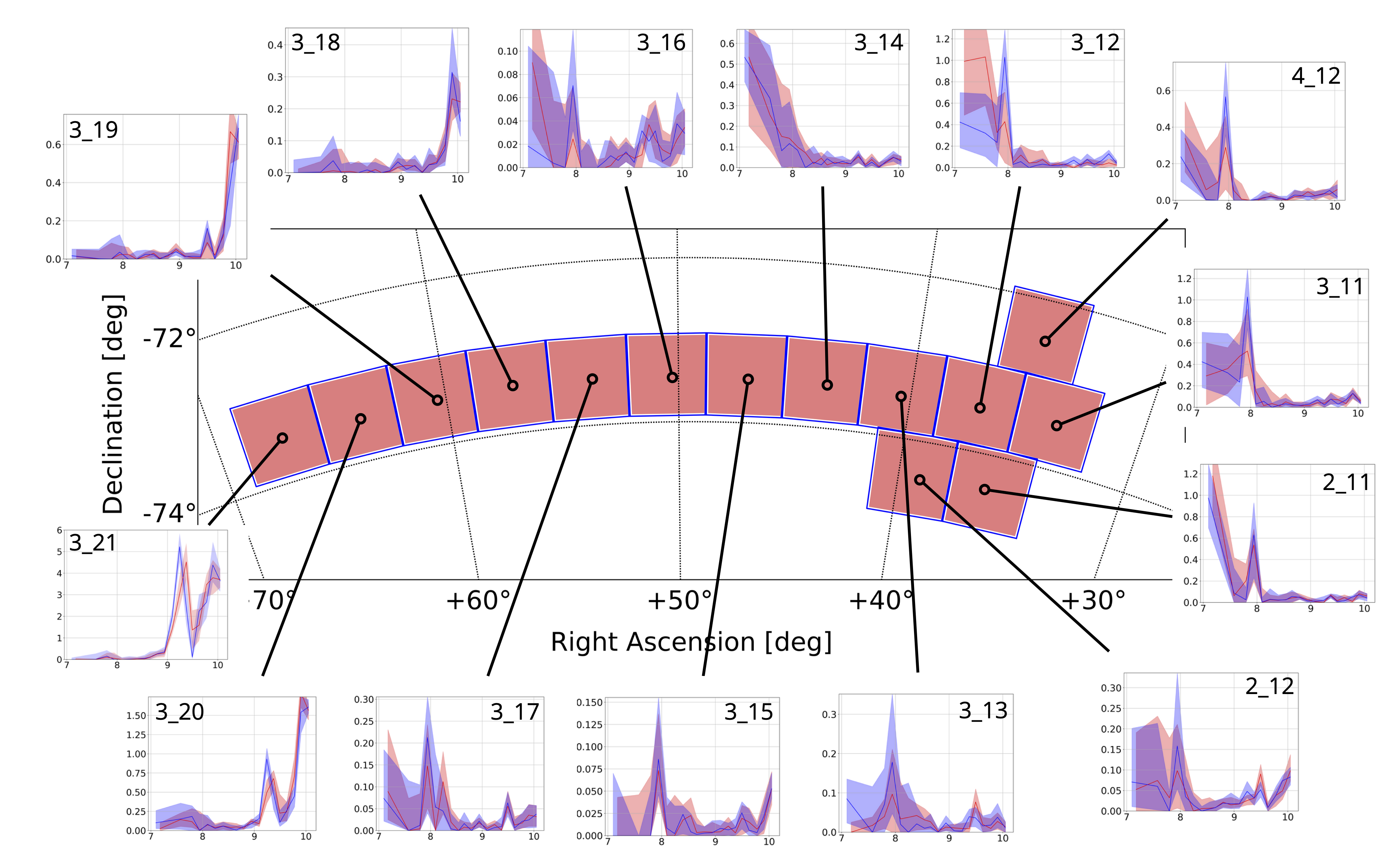}
    \caption{Map of the analyzed regions overlaid with the SFH of the individual tiles. The lines connect the location of each tile on the map to its corresponding inset plot, which displays the SFR in unit of $10^{-4}\, M_{\odot}\,\mathrm{yr}^{-1}$ resulting from the PARSEC-COLIBRI (red) and BaSTI (blue) stellar evolution libraries as a function of age in Gyr (logarithmic scale). The y-axis limits are independently scaled for each tile to better show the SFH features of each region and the axis labels were omitted for enhanced clarity. The shaded regions represent the 5th and 95th percentile uncertainties of the SFR.
}
    \label{fig:mappaSFH}
\end{figure*}

\subsection{The SFH of the Bridge}
\begin{figure}[]
    \centering
    \begin{subfigure}[b]{\textwidth}
        \includegraphics[width=0.47\textwidth,trim={0 15mm 0 55mm},clip]{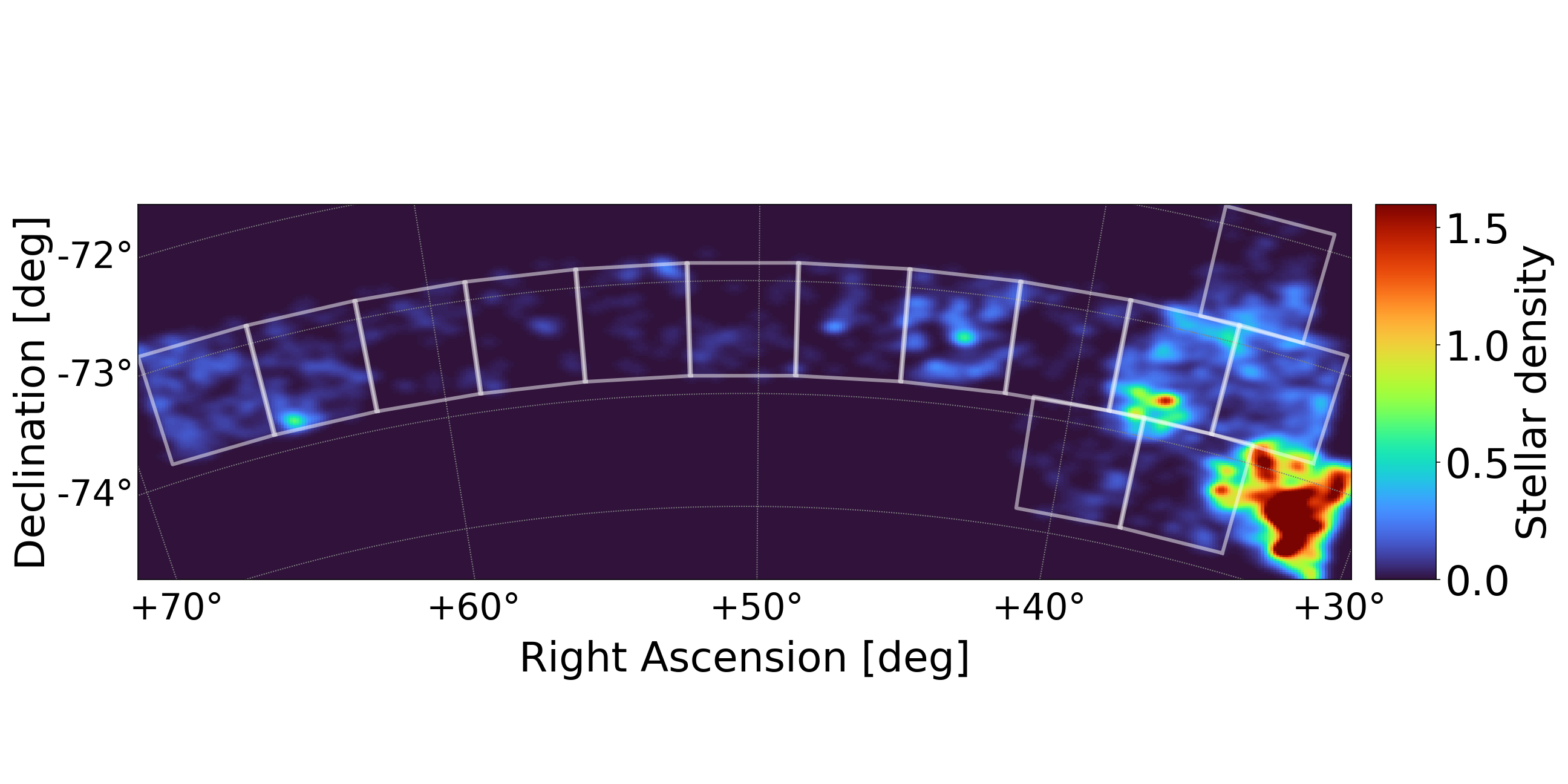}
        \label{fig:UMS}
    \end{subfigure}
    \hfill
    \begin{subfigure}[b]{\textwidth}
        \includegraphics[width=0.47\textwidth,trim={0 15mm 0 55mm},clip]{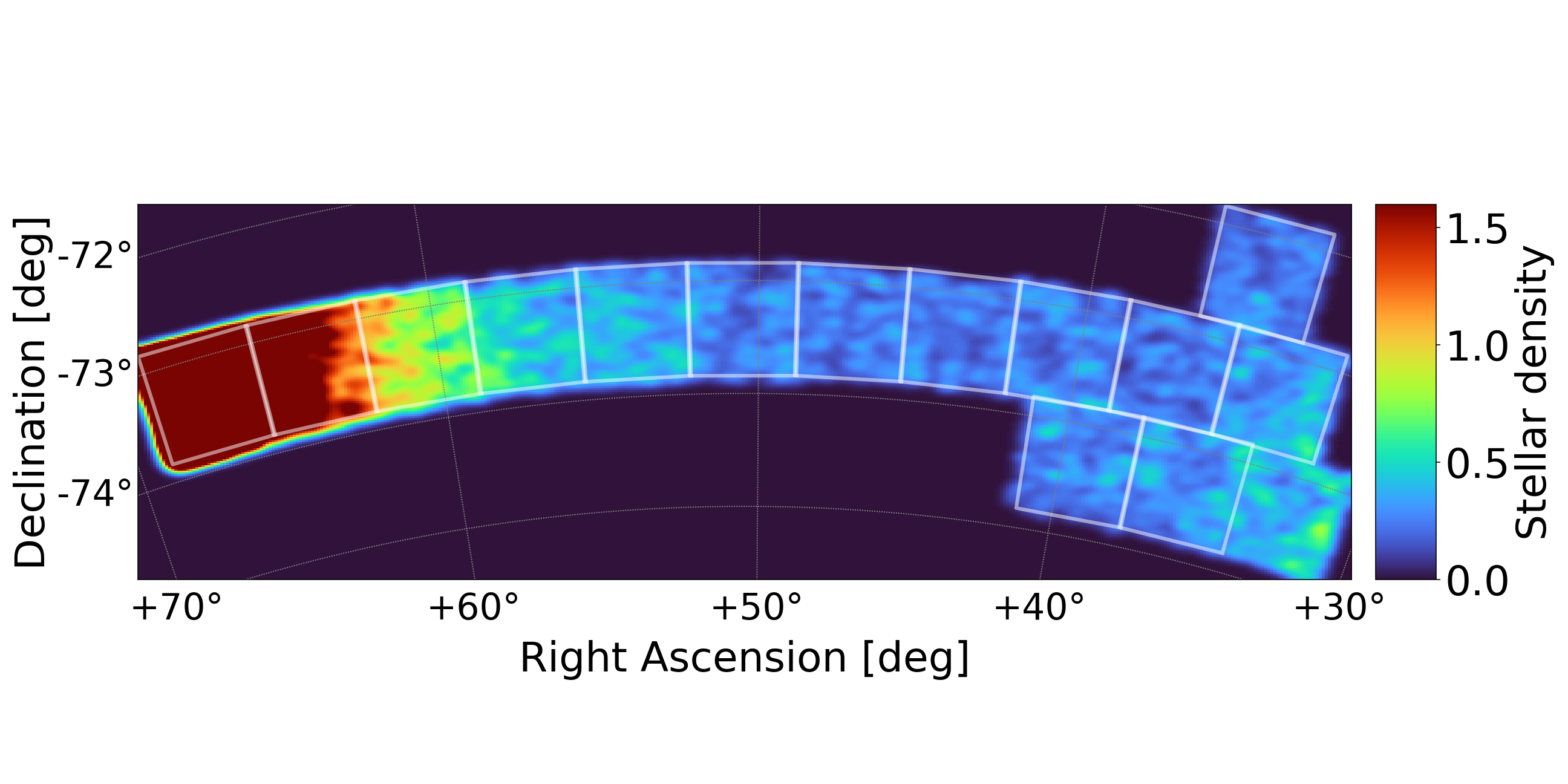}
        \label{fig:RC}
    \end{subfigure}
    \hfill
    \begin{subfigure}[b]{\textwidth}
        \includegraphics[width=0.4705\textwidth,trim={1mm 10mm 0 25mm},clip]{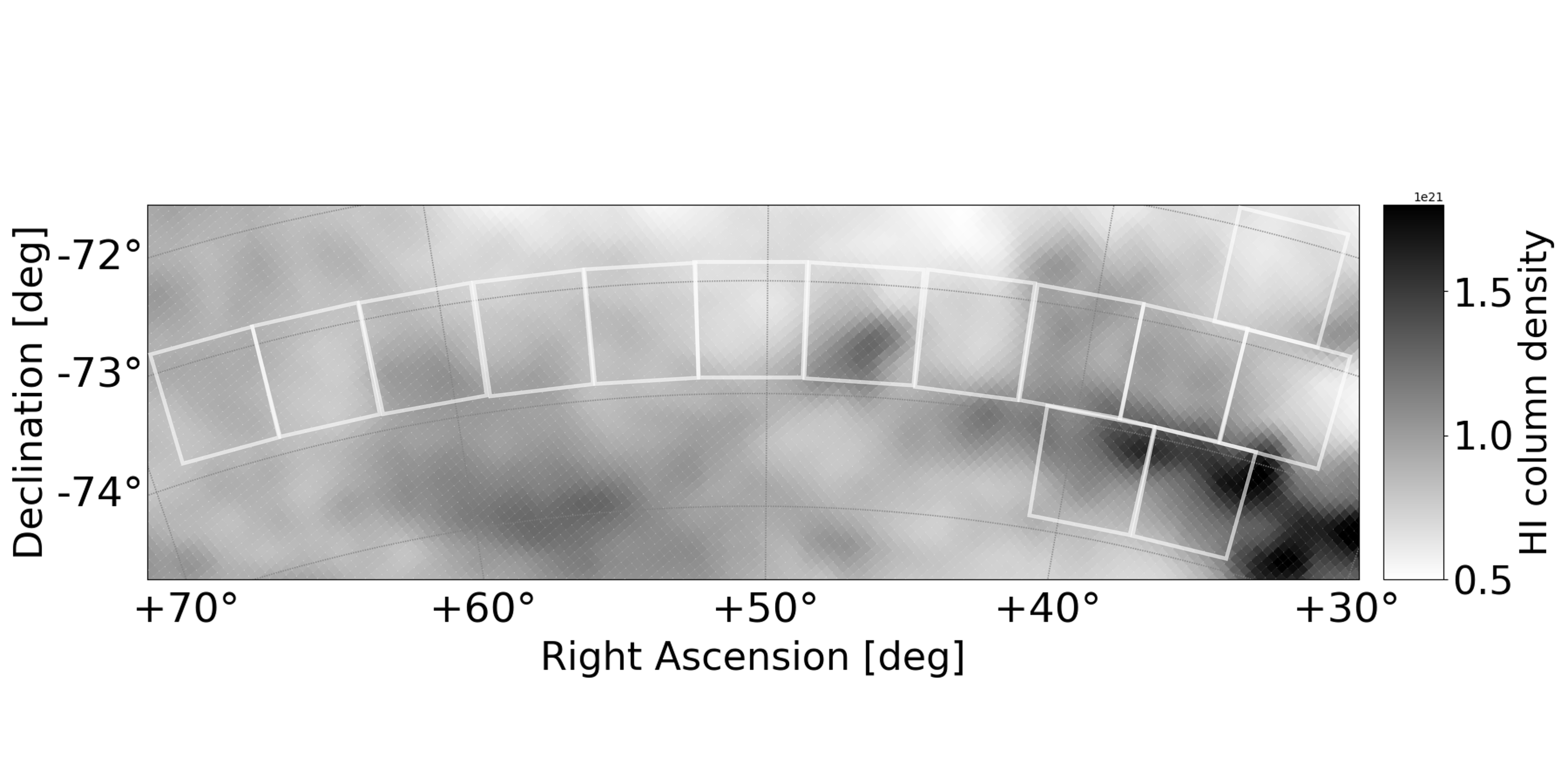}
        \label{fig:HI}
    \end{subfigure}
    \caption{Spatial distribution of upper main sequence stars (top), red giant branch and red clump (centre) and H\,{\small I} gas column density distribution (bottom) from the GASS survey \citep{2009ApJS..181..398M}.}
    \label{fig:3panels_spatialMap}
\end{figure}

Figure \ref{fig:mappaSFH} provides a comprehensive view of the SFH across all the Magellanic Bridge tiles, highlighting significant tile-to-tile variations. Several fields show low recent star formation, while a limited number of tiles, those closer to the SMC, display enhanced activity at ages younger than 150 Myr.
In Figure \ref{fig:3panels_spatialMap} it is clear that upper main sequence stars ($g<20$ and $g-i<-0.2$) are concentrated towards the western side of the Bridge, closer to the SMC. While a broad correspondence between young stellar concentrations and regions of high H\,{\small I} column density is observed in tiles immediately adjacent to the SMC, this correlation becomes weaker within the Bridge itself. \footnote{The young population at RA $\sim3h$ (45°) and $\sim4.3h$ (65°) are massive, short-lived stars ($\lesssim100$ Myr) known from previous studies (e.g.: \citet{1990AJ.....99..191I, 1991A&AS...91..171D, 2021A&A...646A..16R})}.

In particular, tile 3\_14 deserves special attention. Unlike neighbouring fields, its SFH shows a continuous rise up to the present time, resulting in a well-populated young main sequence. This population is clearly visible in its CMD (Figure B1), which provides sufficient statistics to robustly constrain recent star formation. A comparison between the top and bottom panels of Figure \ref{fig:3panels_spatialMap} shows that the region hosting this concentration of young stars coincides with a local depression in the H\,{\small I} column density, suggesting that the gas may have been recently consumed to fuel the ongoing star formation episode. This location corresponds to the 'OGLE Island' reported by \citet{2014ApJ...795..108S}, a small isolated young population in the Magellanic Bridge. In contrast, the adjacent tiles contain a larger amount of gas but show fewer upper main sequence stars, as evidenced in Fig. \ref{fig:CMDs_tiles}, suggesting an inhomogeneous star-forming efficiency across small spatial scales in the Bridge.
This absence of very young stars beyond RA = $03h$ (45°) was already noted by \citet{2007ApJ...658..345H}, who suggested that it might be linked to the drop in H\,{\small I} surface density below the threshold required to trigger SF \citep{1989ApJ...344..685K}.

The central tiles of the Bridge (3\_15, 3\_16 and 3\_17) do not show evidence for a stellar population younger than $\sim400$ Myr. This is reflected both in the corresponding SFHs (note that the maximum y-axis value is significantly lower than in the adjacent regions) and in the paucity of UMS stars in the density map (upper panel of Figure \ref{fig:3panels_spatialMap}). A visual inspection of the CMDs (Figure \ref{fig:CMDs_tiles}) further supports this conclusion, as no distinct young main sequence comparable to that observed in tile 3\_14 is detected. Together, these indicators suggest that recent star formation  in the central Bridge is extremely weak, if present at all.

Assuming a tangential velocity of approximately $\sim80-110$ km/s \citep{2018ApJ...864...55Z, 2020A&A...641A.134S, 2021A&A...649A...7G}, the young stellar population of the Bridge likely formed in situ or in the nearby regions of the SMC. Given such a velocity, it would not have had sufficient time to migrate more than $\sim8$ kpc over such a short timescale. This is consistent with the lower H\,{\small I} gas velocity observed in the Wing, traced by the young, massive stars embedded within it \citep{2019ApJ...887..267M}. 

Concerning older populations, we detect a modest amount of ancient ($\gtrsim8$ Gyr) stars in the inter-cloud region, which appear to be uniformly distributed along the Bridge and likely pertains to the halos of the two Clouds. However, in the central parts of the Bridge, the CMDs are dominated by the contamination from Milky Way stars. The absence of clumps in the distribution of Red Clump and Red Giant Branch stars ($19.5 \lesssim g \lesssim 22$ mag and $0.6\lesssim g-i \lesssim 1.2$ mag, in the central panel of Figure \ref{fig:3panels_spatialMap}) may imply that the primary outcome of the last pericentric passage was the stripping of gas rather than stars. 

In tiles 3\_19, 3\_20, and 3\_21, dominated by the outer LMC population, the SFH shows a dominant old population ($>7$ Gyr) as well as a distinct peak at $\sim2$ Gyr. The former is consistent with the oldest star formation seen in \citet{2007ApJ...658..345H} close to the LMC (region mb20) and in \citet{2020A&A...639L...3R}, while the latter corresponds to a major star formation episode found in the global SFH of both Clouds \citep{2022MNRAS.513L..40M}.

\begin{figure}[h]
    \centering
    \includegraphics[width=0.48\textwidth,trim={0mm 0mm 0mm 0mm},clip]{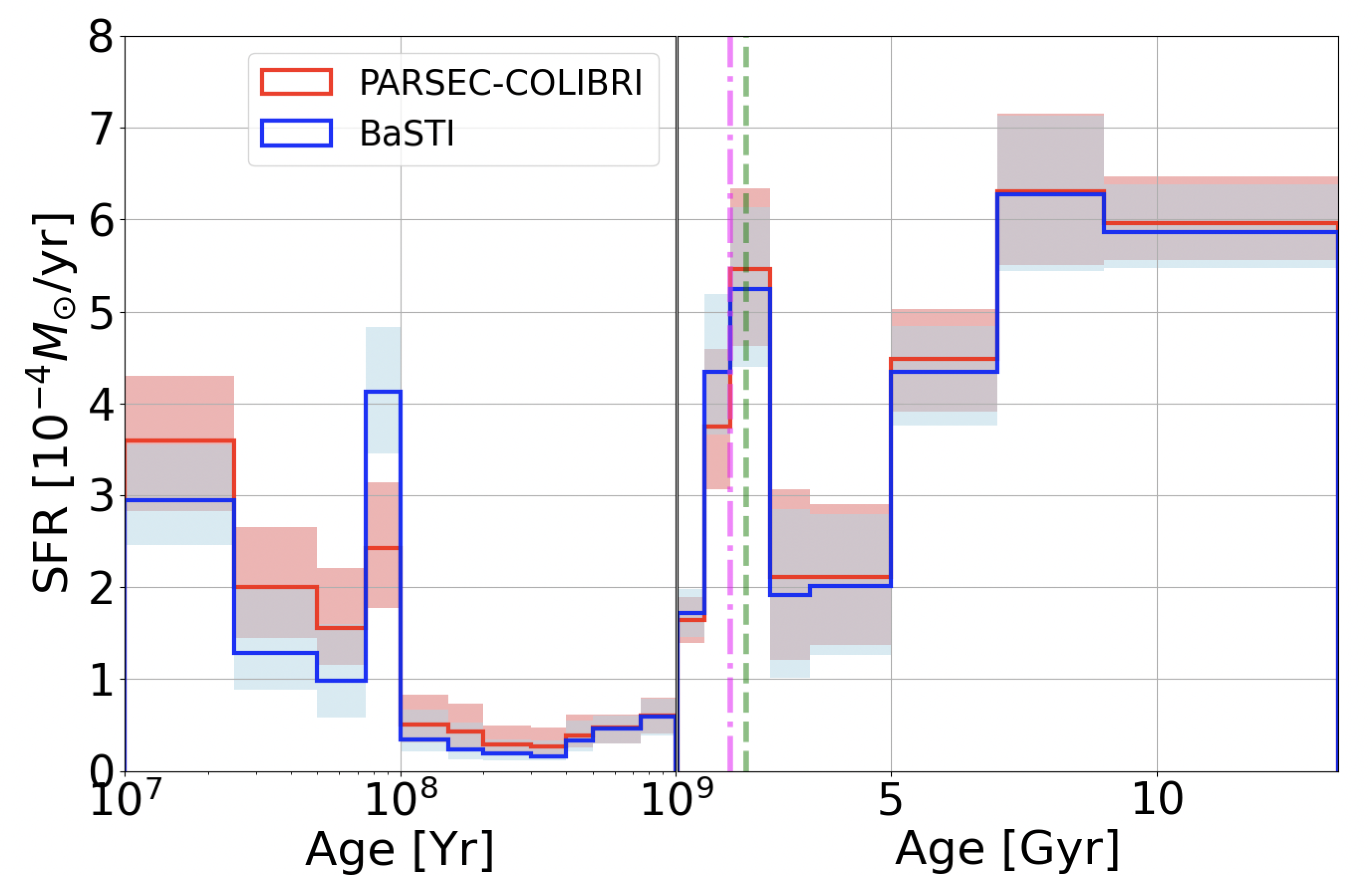}
    \caption{Comparison between the global SFH of the Magellanic Bridge recovered using the PARSEC-COLIBRI (red line) and BaSTI (blue line) stellar models. The pink dashed line indicate a period of intense SF found by \citet{2021MNRAS.508..245M}; the green dashed line indicate the first synchronized SF peak of the MCs found by \citep{2022MNRAS.513L..40M} and \citep{2020A&A...639L...3R}.
    Notice that the left portion of the abscissa is in logarithmic age scale to better appreciate the details, while the right portion is in linear scale. The shaded area represents the confidence interval of our solution, computed as the \(5^{th}\) and \(95^{th}\) percentiles of the parameters’ distribution.}
    \label{fig:tot_SFR}
\end{figure}

Figure \ref{fig:tot_SFR} presents the global SFH of the Bridge, obtained by summing the SFHs of the individual tiles, for both stellar models adopted. The dashed lines indicate periods of enhanced SF activity found by \citet{2020A&A...639L...3R}, \citet{2021MNRAS.508..245M} and \citet{2022MNRAS.513L..40M}. Given that the SFR uncertainties in each tile are independent, the total error was calculated as the square root of the sum of the squared individual errors. We found that the Magellanic Bridge hosts ongoing SF, which primarily began approximately 100 Myr ago. The main difference between the two stellar models lies in the amount of SF from 100 Myr ago to the present day, which may reflect differences in how young stellar populations are treated by the two sets of stellar evolution codes and their high-mass limit. For all other epochs, the SFHs derived from the two sets show excellent agreement. The old periods of enhanced SF at $\sim2$ Gyr and $\sim10$ Gyr are consistent with the main features of activity observed in previous works in the literature. Interestingly, the peak at $\sim2$ Gyr may suggest a previous close encounter between the two Clouds at that time, as a similar feature can be found in previous works in the literature in both the SMC and LMC \citep{2020A&A...639L...3R, 2021MNRAS.508..245M, 2022MNRAS.513L..40M}. Recent works \citep{2020AJ....159...82B,2023MNRAS.524.2244O} on the clusters of the Bridge have found different populations: an older, metal-poor group ($0.5-6.8$ Gyr, $[$Fe/H$]<-0.6$ dex) likely stripped from the SMC outskirts, and a younger population ($<200$ Myr) with significantly higher metallicities ($-0.5<[$Fe/H$]<-0.1$), which likely formed in situ. The well-established peak in the age distribution of star clusters around 2–3 Gyr in both galaxies \citep{2005AJ....129.2701R, 2016MNRAS.461..519P, Gatto2020, 2024A&A...690A.164G, 2024MNRAS.528.2274D} provides additional evidence that such an event could have triggered enhanced SF across the Magellanic system. This period likely coincides with the formation epoch of the Magellanic Stream and the arrival of the Clouds into the Milky Way halo \citep{2012MNRAS.421.2109B, 2020Natur.585..203L}.
We estimated the total stellar mass in the Bridge, excluding tiles 3\_21, 3\_20, 3\_19 and 3\_18, where the population is almost entirely composed of old LMC stars. For each tile, we calculated the mass from the best-fit solution of each bootstrap run and then took the median of the 25 resulting estimates. Summing over all tiles, we obtain a total mass of $M_* \sim (5.1 \pm 0.2)\times10^{5}$ $M_\odot$, which lies at the upper end of the of minimum stellar mass range proposed by \citet{2023MNRAS.524.2244O}, that is already significantly higher than previous estimates \citep{2007ApJ...658..345H}.

\subsection{Age-metallicity relation}

\begin{figure}[]
    \centering
    \includegraphics[width=0.5\textwidth,trim={10mm 0 0 10mm},clip]{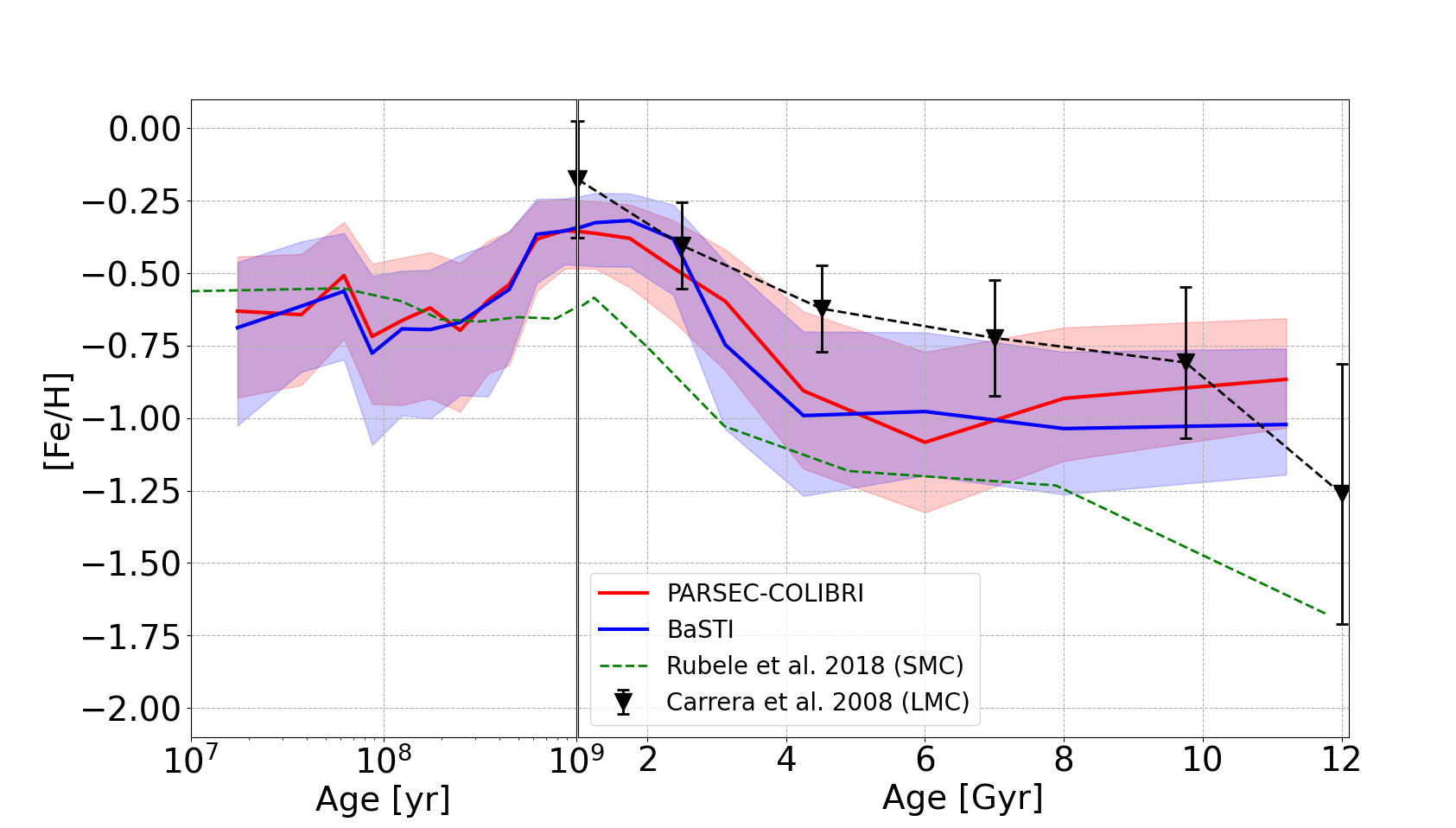}
    \caption{Global AMR of the Magellanic Bridge found using the PADOVA-COLIBRI (red line) and BaSTI (blue line) stellar models. The left portion of the abscissa is in logarithmic scale to better appreciate the details, while the right portion is in linear scale. The shaded area represents the confidence interval of our solution, computed as the 5th and 95th percentiles of the parameters’ distribution. The green dashed line marks the global AMR of the SMC obtained by \citet{2018MNRAS.478.5017R}, while the black dashed line represents the AMR of the LMC from \citet{2008AJ....135..836C}.}
    \label{fig:AMR}
\end{figure}

Regarding the AMR, we did not place any constraints on the SFH recovery algorithm, allowing the metal-content to vary in the range $-2<[$Fe/H$]<0$ dex. Figure \ref{fig:AMR} displays the best fits of the AMR for the entire Magellanic Bridge. The metallicities from equation \eqref{eq:Z} were averaged for each tile. Each tile’s contribution was weighted by the number of stars above a threshold magnitude, to account for differences in completeness. We highlight that the AMR features at older ages ( $>1$ Gyr) mainly reflect the tiles near the LMC, while the younger populations are dominated by tiles closer to the SMC. Therefore, the overall AMR is not the result of a coherent chemical evolution, but rather reflects the mixing of stellar populations originating from different parent systems. Only the youngest stars are likely to have formed in situ within the Bridge. The results from both adopted stellar libraries indicate that the metallicity slowly grew up to approximately 5 Gyr ago, after which the production of metals began to accelerate, up to $\sim2$ Gyr ago. The metallicity values we recovered for stars older than $1$ Gyr are systematically lower than those of \citet{2008AJ....135..836C}, and this discrepancy may be attributed to the fact that they analyzed the disk of the LMC, and did not reach the periphery of the galaxy. 

Although metallicity gradients in dwarf galaxies were once debated, improved observations have now revealed their presence in many systems. Early investigations that did not detect such gradients \citep[e.g., ][]{2006ApJ...642..813L, 2009ApJ...700..309B, 2009ApJ...707.1676C} were likely limited by the restricted spatial coverage and sensitivity of available data. Several studies have indeed reported declining metallicity with radius \citep{2014MNRAS.442.1680D, 2015MNRAS.450.3254P, 2015AJ....150..143A,  2017ApJ...843...20A, 2019MNRAS.482.3892A, 2021MNRAS.507.4752C,2022A&A...665A..92T,2024AJ....167..123L}. Recent studies based on the analysis of a large number of high resolution spectra have clearly detected metallicity gradients in both the LMC and SMC \citep{2025MNRAS.544..457P, 2025MNRAS.544..430P, 2025A&A...700A..74O}, and the lower metallicities found in this work could be influenced by these radial trends. 
The plot shows that the AMR lies between the typical SMC and LMC values, and broadly follows the expected spatial trend from the SMC side toward the LMC along the Bridge. The metallicity decreases toward SMC-side regions of the Bridge; however, the uncertainties are larger at these epochs because of the generally low star-formation activity in these regions. In addition, the metallicities derived from photometry are subject to the well-known age–metallicity degeneracy, which further increases the uncertainty.

We found that the young stellar population (left portion of Figure \ref{fig:AMR}) have $[$Fe/H$]\sim -0.6$ dex, similar to the results of \citet{2018MNRAS.478.5017R} in the SMC, suggesting that the material forming the young population of the Bridge was stripped primarily from the SMC during the last close encounter between the Clouds. Interestingly, a metallicity dip has also been reported for clusters around $\sim200$ Myr ago, with metallicity decreasing from $-0.25$ to $-0.55$ dex \citep{2023MNRAS.524.2244O}, suggesting a second infall of metal-poor gas. The stellar metallicity we observe is higher than that measured for three O-type stars in the eastern outskirts of the SMC \citep{2021A&A...646A..16R}. This discrepancy could arise due to inefficient mixing from stellar winds and supernovae in the Bridge, combined with additional chemical inhomogeneities caused by the accretion of metal-poor gas by interactions between the MCs.

\begin{figure}
    \centering
    \includegraphics[width=0.47\textwidth, trim={15mm 10mm 40mm 30mm},clip]{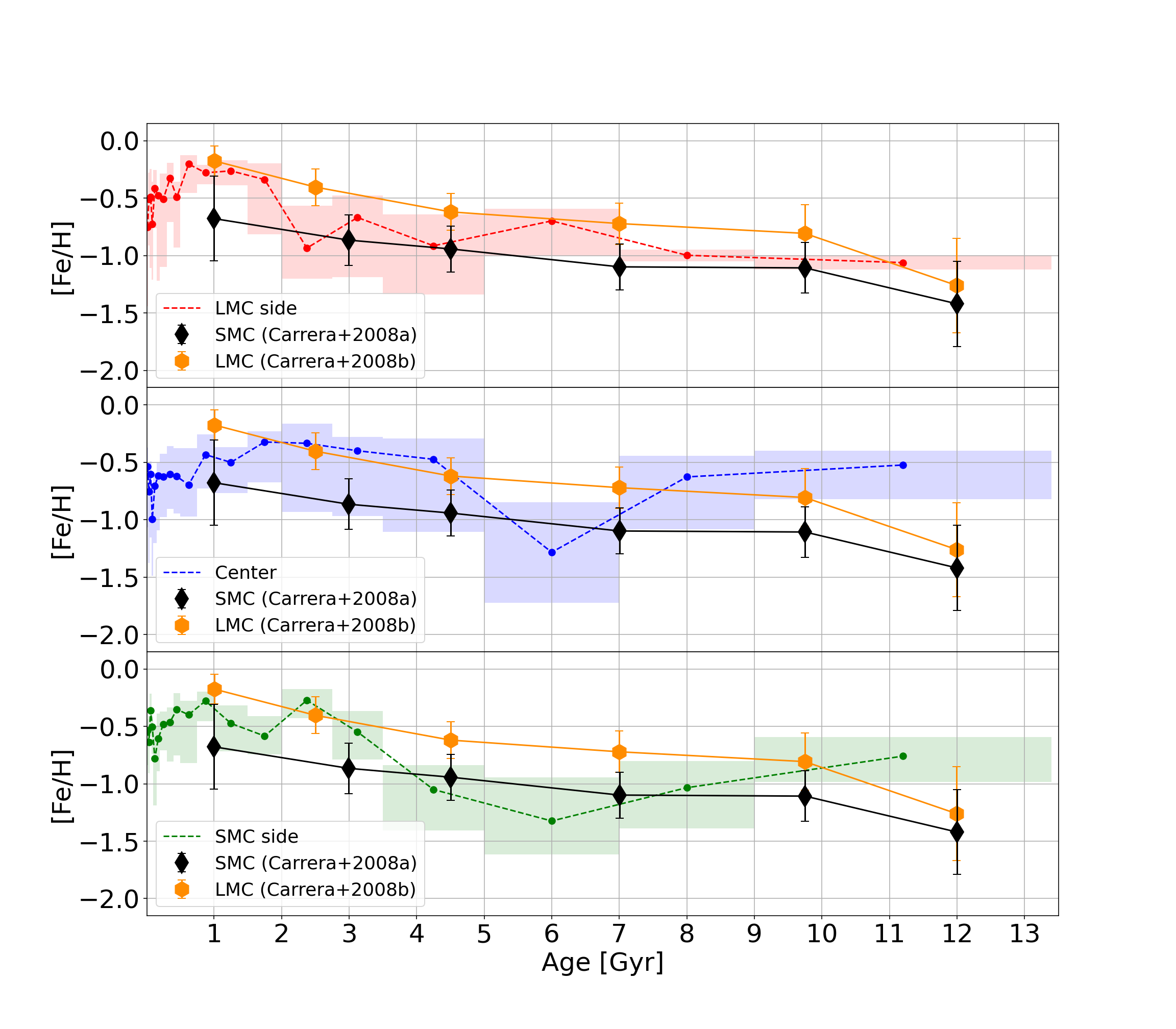}
    \caption{AMR for the three observed regions, with the top, central, and bottom panels representing the LMC-side, central, and SMC-side regions, respectively. Each point corresponds to the mean metallicity within an age bin, with horizontal error bars showing the bin width. Black diamonds and orange hexagons indicate spectroscopic measurements for the SMC \citep{2008AJ....136.1039C} and LMC \citep{2008AJ....135..836C}, respectively.}
    \label{fig:AMR_reg}
\end{figure}

\subsection{Star formation efficiency in the Bridge: testing the Kennicutt–Schmidt relation}

To investigate the efficiency of star formation in the Bridge, we compared the observed gas and star formation rate surface densities. This allows us to test whether the Bridge follows the same scaling laws observed in other galaxies, or whether its low density, low metallicity, and tidally disturbed environment lead to a reduced efficiency in forming stars. The key empirical relationship describing the conversion of gas into stars is the Kennicutt-Schmidt (KS) law \citep{1989ApJ...344..685K, 1998ApJ...498..541K}, which expresses a power-law correlation between the SFR surface density ($\Sigma_{\text{SFR}}$) and the gas mass surface density ($\Sigma_{\text{gas}}$) in galaxies:
\begin{equation}
    \Sigma_{\text{SFR}} \propto (\Sigma_{\text{gas}})^n
    \label{eq:ks_proportionality}
\end{equation}
where $n \sim 1.0 - 1.4$ for a wide range of galaxy types \citep{1989ApJ...344..685K,2019ApJ...872...16D,2023ApJ...945L..19S}. The classical KS relation tends to hold for massive galaxies, but often deviates for gas-rich lower-mass dwarf galaxies, especially metal-poor ones like the MCs. This deviation depends on the dominant gas phase (atomic vs. molecular), the galactic environment, and the spatial scales considered. The \citet{2009ApJ...699..850K} model (KMT09) refines the KS relation as:
\begin{equation}
    \Sigma_{\text{SFR}} = \frac{f_{\text{H}_2} \Sigma_{\text{gas}}}{t_{\text{dep}}}
    \label{eq:sfr_molecular_gas}
\end{equation}
where $t_{\text{dep}}$ is the molecular gas depletion time and $f_{\text{H}_2}$ is the molecular gas fraction, determined by the balance between the dissociation and the formation of molecules on dust grains' surfaces. In the $f_{\text{H}_{2}}(\Sigma_{\text{gas}},c,Z')$ parameter, KMT09 accounts for the gas metallicity (Z') and the degree of gas clumping (c). Metallicity affects dust abundance, which in turn influences H$_2$ formation and far-ultraviolet (FUV) radiation field shielding. Gas clumping, on the other hand, directly impacts local density and shielding. The dependence on these factors leads to a characteristic S-shaped curve with an atomic-dominated regime for low $\Sigma_{\text{gas}}$, where $\Sigma_{\text{SFR}}$ is significantly suppressed and deviates strongly from the classical KS relation. Therefore, KMT09 provides a more physically motivated description of the star formation law, explaining why simple power-law relations often fail to capture the observed variations in diverse galactic environments.

\begin{figure}[htb!]
    \centering
    \includegraphics[width=0.44\textwidth]{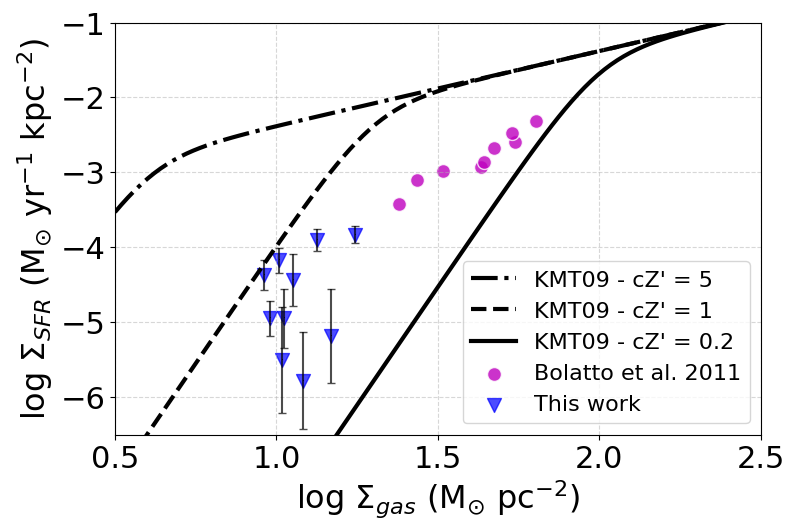}
    \caption{Star formation law in the Magellanic Bridge (blue triangles) in comparison with the data by \citet{2011ApJ...741...12B} (magenta filled circles) and curves of the KMT09 models (adapted from the same authors) for different values of the $cZ'$ parameter (the product of the gas metallicity $Z'$ and the degree of gas clumping $c$) indicated by the dash-dotted, dashed, and solid lines.
    }
    \label{fig:KS}
\end{figure}

To determine $\Sigma_{\text{gas}}$, we utilized the HI4PI survey data \citep{2016A&A...594A.116H}. The HI4PI survey provides all-sky maps of H\,{\small I} column density ($N_{\text{HI}}$) in units of cm$^{-2}$. We averaged the individual $N_{\text{HI}}$ values within each of the studied regions and converted the average H\,{\small I} column density into a surface density $\Sigma_{\text{HI}}$:
\begin{equation}
    \Sigma_{\text{HI}} = N_{\text{HI}} \times m_{\text{H}} \times \frac{1 \ \text{M}_{\odot}}{1.989 \times 10^{33} \ \text{g}} \times \frac{9.523 \times 10^{36} \ \text{cm}^2}{1 \ \text{pc}^2}
    \label{eq:hi_surface_density}
\end{equation}
where $m_{\text{H}} = 1.673 \times 10^{-24}$ g is the mass of a hydrogen atom. This simplifies to approximately $N_{\text{HI}} \times 8.02 \times 10^{-21}$ M$_{\odot}$ pc$^{-2}$. In absence of molecular hydrogen (H$_2$) column density maps in the literature, we assumed a fixed 10\% contribution from H$_2$ to the total gas mass, consistent with values for dwarf galaxies and outer galactic disks where H\,{\small I} often dominates \citep{2011ApJ...741...12B, 2019ApJ...885L..32D} and the molecular fraction is typically low ($\sim10-20\%$). Finally, we compared the gas column density with $\Sigma_{\text{SFR}}$, that we calculated as the SFR in the most recent age bin of each tile divided by the tile area ($\sim1$ kpc). As shown in Figure \ref{fig:KS}, $\Sigma_{\text{SFR}}$ follow the slope expected in the KMT09 theoretical framework. We only show the results from the PARSEC-COLIBRI library for clarity, the results with the BaSTI models are similar. Our study extends this established trend to even smaller values of $\Sigma_{\text{HI}}$, in a regime where molecular gas formation is less efficient. The observed trend in our data aligns well with the trend observed by \citet{2011ApJ...741...12B}, who examined the star formation efficiency in the SMC. 

\section{Conclusions}
In this work, we analyzed the SFH of the entire Magellanic Bridge, covering a total area of 14 deg² using data from the STEP Survey. We employed the ‘synthetic CMD’ method and derived the best-fitting SFH using the genetic algorithm SFERA. Our library of partial models covered the whole Hubble time and a metallicity range of $-2.0\leq[$Fe/H$]\leq0$. To explore systematic differences, we used two distinct sets of stellar evolutionary models: PARSEC-COLIBRI and BaSTI.
For each analyzed tile, we assumed a distance based on a linear trend between the distances of the LMC and the SMC. Extinction corrections were derived from the SFD extinction map. Thanks to the high photometric accuracy of the STEP survey, we were able to map the spatial distribution of stellar ages and trace the metallicity evolution across the Magellanic Bridge with high temporal resolution. Our findings significantly extend previous analyses, covering a region more than three times larger than the most comprehensive SFH study of the Bridge by \citet{2007ApJ...658..345H}. The key results include:

\begin{itemize}
\item The young stellar population is more evident in the region of the Magellanic Bridge closer to the SMC. The solutions from both PARSEC-COLIBRI and BaSTI models consistently predict low star formation activity before $\sim100$ Myr ago, when star formation was ignited and continued to rise up to the present day.  This timeframe aligns with predictions from dynamical models for the Magellanic Bridge's formation, and the metallicity of stars from this event matches that of the SMC.

\item A very old stellar population originated in the LMC is observed in the three tiles closer to the LMC. These periods of enhanced star formation are consistent with peaks found in previous works in the LMC ($\sim10$ Gyr) and in both MCs ($\sim2.1$ Gyr) \citep{2018MNRAS.478.5017R, 2020A&A...639L...3R, 2021MNRAS.508..245M}

\item We observe only minimal differences between the SFHs recovered using the PARSEC-COLIBRI and BaSTI stellar libraries, except for the height of the star formation rate (SFR) peak at 100 Myr, which arises from different physical assumptions in the PARSEC-COLIBRI and BaSTI stellar models. These differences are most pronounced for the young, intermediate-mass stars that dominate a 100 Myr population and thus are probably related to the treatment of core convective overshooting, mass loss, and post-MS evolution. This leads to model-dependent stellar luminosities, which in turn affects the inferred number of stars required to match the observational data.

\item The metallicity evolution at old ages follows the trend observed in the LMC, though it is systematically lower. This may be due to the fact that we are observing material originating from the outer regions of the LMC, as metallicity is expected to decrease with galactocentric radius. In the past 400 Myr, the AMR closely matches the global AMR of the SMC, further supporting the scenario that most of the material forming the Bridge was stripped from the SMC.

\item The present-day metallicity is $[$Fe/H$]\sim-0.6$ dex, higher than that derived for three O-type stars in the SMC's eastern outskirts \citep{2021A&A...646A..16R}. This may be due to chemical inhomogeneities caused by previous interactions between the MCs, as well as inefficient mixing processes from stellar winds and supernovae within the Bridge.

\item We estimated the total stellar mass formed in the Magellanic Bridge, excluding those dominated by old LMC stars. The resulting total mass across $\sim10$ deg$^2$ is $M_* \sim (5.1 \pm 0.2)\cdot10^{5}$ $M_\odot$, higher than previous estimates \citep{2007ApJ...658..345H, 2023MNRAS.524.2244O}.

\item We used HI4PI survey data to estimate the $\Sigma_{\text{gas}}$ in the Bridge and compared it with our resulting $\Sigma_{\text{SFR}}$. We see good agreement with the theoretical framework of the KMT09 model, extending the trend observed by \citet{2011ApJ...741...12B} to a regime where the formation of molecular gas, and therefore star formation, is less efficient.
\end{itemize}
Our findings suggest the recent collision between the SMC and LMC likely occurred no more than a few tens of Myr before the peak we found at 100 Myr ago. This coincides with the predicted time when the SMC crossed the LMC’s disk plane \citep{2013MNRAS.428.2342B,2022ApJ...927..153C}. Our results can help place constraints on models of the MCs’ interaction history and better understand the tidal origin of the stellar populations in the Bridge.

\begin{acknowledgements}
We thank our anonymous referee for the insightful comments and suggestions that helped us improve the paper. This research has used the SIMBAD database operated at CDS, Strasbourg, France. We acknowledge funding from: INAF GO-GTO grant 2023 “C-MetaLL - Cepheid metallicity in the Leavitt law” (P.I. V. Ripepi); PRIN MUR 2022 project (code 2022ARWP9C) 'Early Formation and Evolution of Bulge and Halo (EFEBHO),' PI: Marconi, M., funded by the European Union – Next Generation EU; Large Grant INAF 2023 MOVIE (P.I. M. Marconi). 
This research has made use of the APASS database, located at the AAVSO web site. Funding for APASS has been provided by the Robert Martin Ayers Sciences Fund.
M.B. acknowledges the financial support by the Italian MUR through the grant PRIN 2022LLP8TK\_001 assigned to the project LEGO -- Reconstructing the building blocks of the Galaxy by chemical tagging (P.I. A. Mucciarelli), funded by the European Union -- NextGenerationEU.
M.G. acknowledges "Partecipazione LSST - Large Synoptic Survey Telescope (ref. A. Fontana)" (Ob. Fu.: 1.05.03.06).

\end{acknowledgements}

%
%

\bibliographystyle{aa}
\bibliography{name.bib}

\appendix
\section{Comparison of MW Contamination Models}
\label{sec:MW_contamination}

To further illustrate the impact of different MW contamination modeling approaches, we present here a comparison of the SFH recovery across different tiles using both the YMCA tile and synthetic CMDs of the MW generated by TRILEGAL. We generated synthetic foreground CMDs for fields centered on the analyzed tiles, each covering 1 deg\(^2\). The synthetic foreground CMD was then convolved with the incompleteness of the data and incorporated into our library of partial CMDs.

\begin{figure}[]
    \centering
    \includegraphics[width=0.5\textwidth, trim={0 0 20mm 10mm},clip]{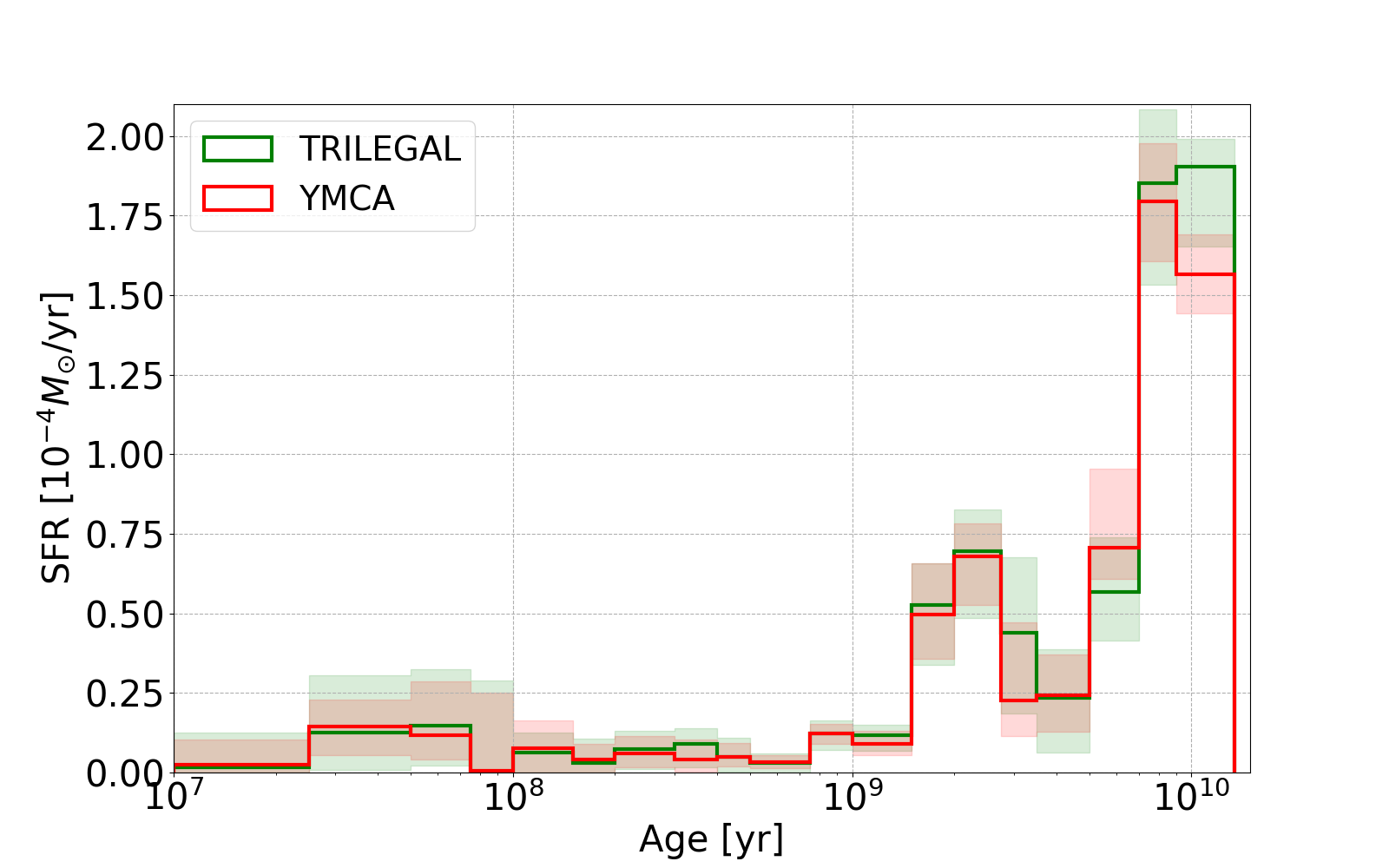}
    \caption{Comparison of the recovered SFRs of tile 3\_20 using the two different models for MW foreground contamination. The red histogram shows the SFR derived using the observed YMCA tile as foreground model, while the green histogram is the result with the TRILEGAL synthetic CMD.}
    \label{fig:320_TRI_YMCA}
\end{figure}

\begin{figure}[]
    \centering
    \includegraphics[width=0.5\textwidth, trim={0 0 20mm 10mm},clip]{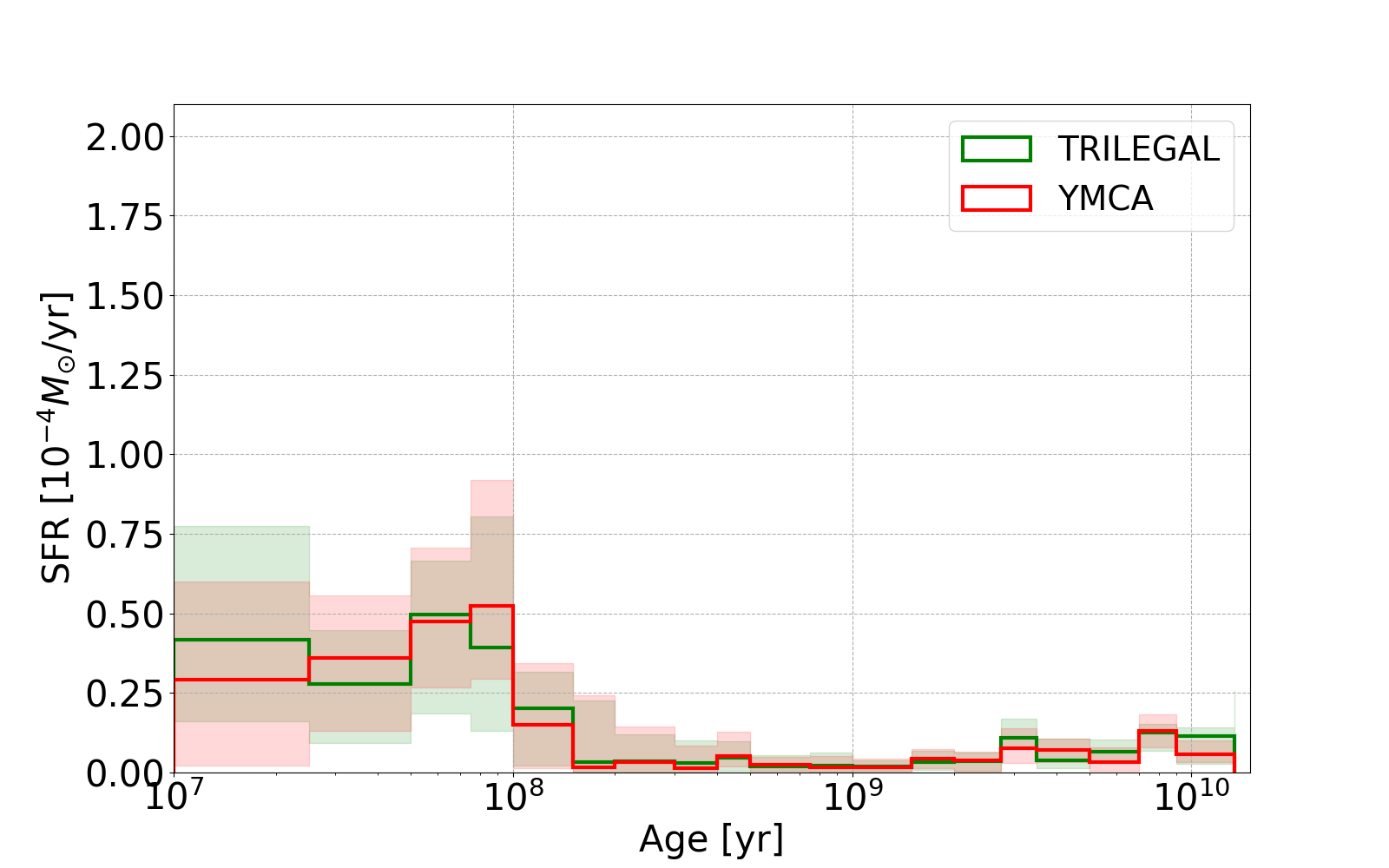}
    \caption{Same as Figure \ref{fig:320_TRI_YMCA} for tile 3\_11.}
    \label{fig:311_TRI_YMCA}
\end{figure}

\begin{figure}[]
    \centering
    \vbox{
        \includegraphics[width=0.35\textwidth, trim={0 0 0 30mm},clip]{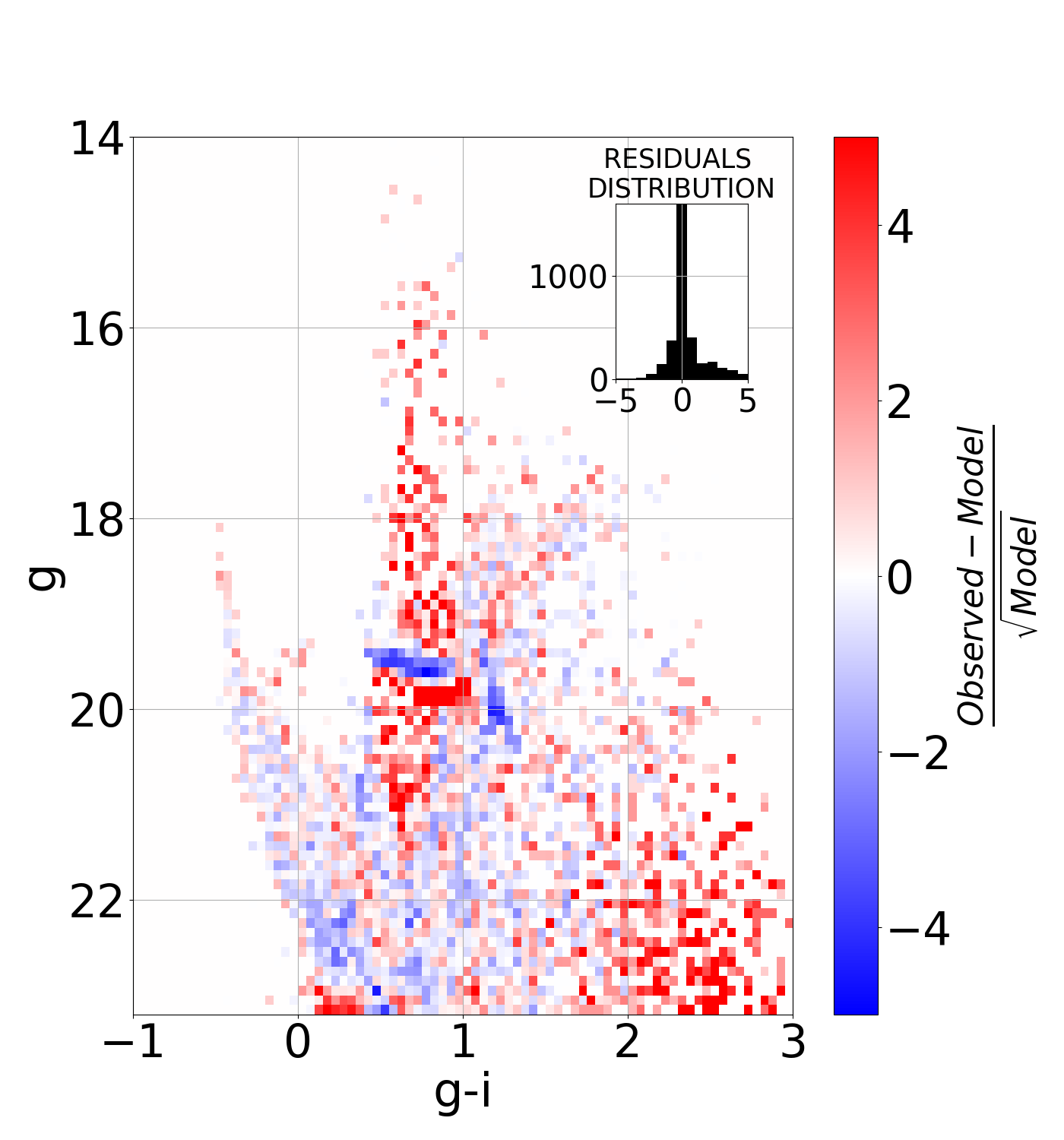}
        \vspace{-0.1cm}
        \includegraphics[width=0.35\textwidth, trim={0 0 0 30mm},clip]{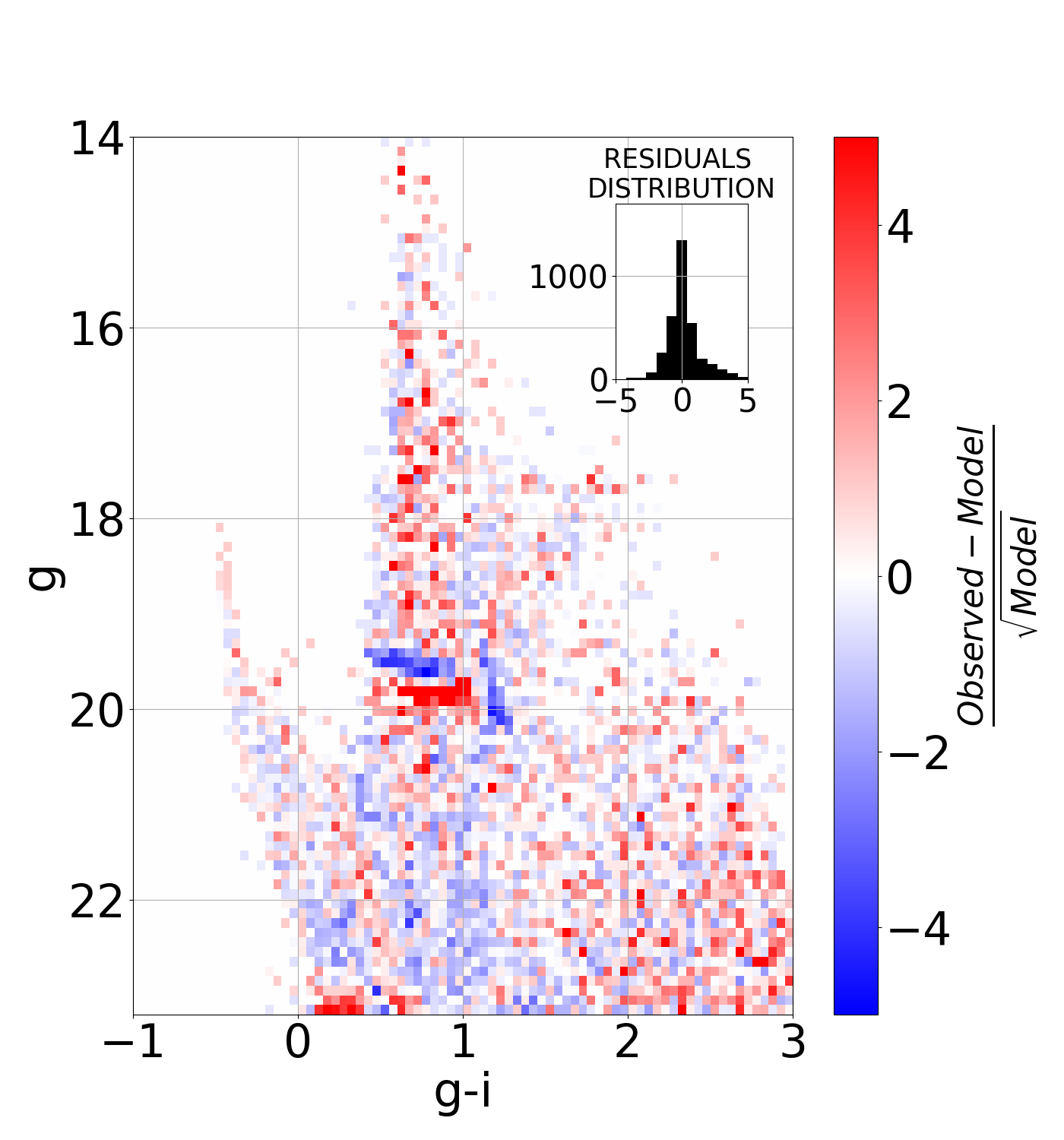}}
    \caption{Comparison of the residuals' Hess diagrams for tile 3\_20 using TRILEGAL (top) and the observed YMCA tile (bottom) as foreground model.}
    \label{fig:confronto320}
\end{figure}

\begin{figure}[]
    \centering
    \vbox{
        \includegraphics[width=0.35\textwidth, trim={0 0 0 30mm},clip]{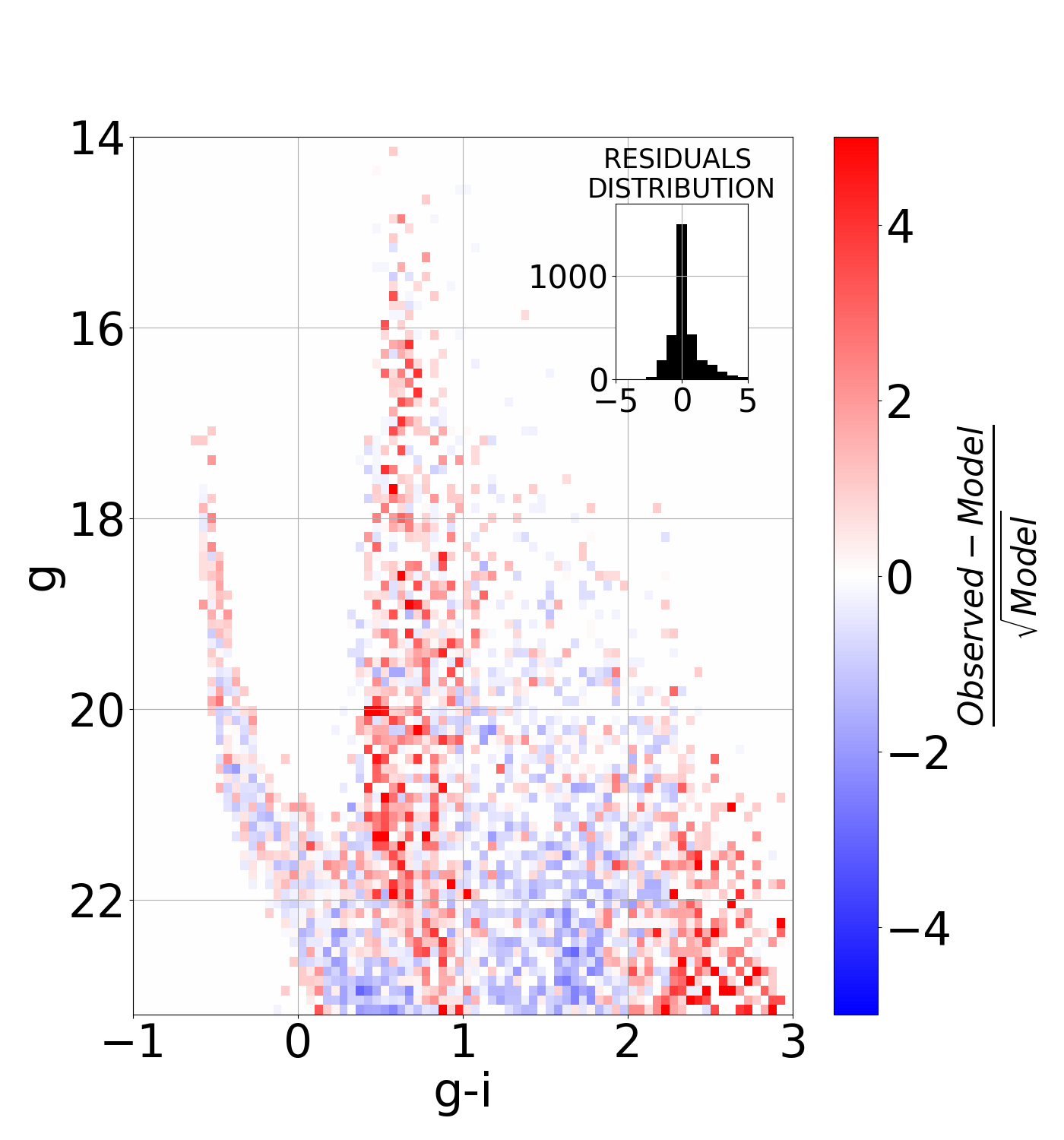}
        \label{fig:sub1}
        \vspace{-0.1cm}
        \includegraphics[width=0.35\textwidth, trim={0 0 0 30mm},clip]{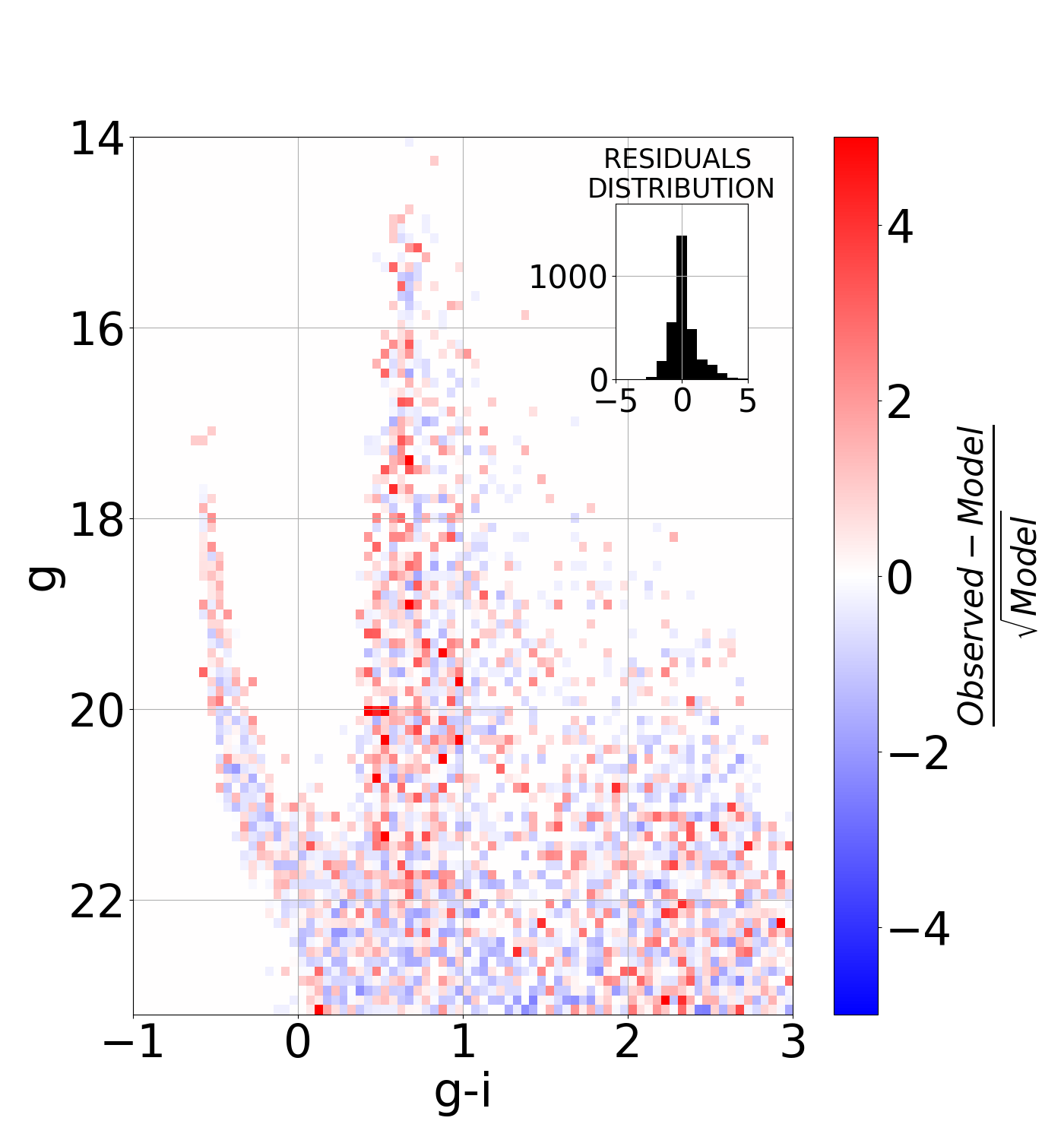}}

        \label{fig:sub2}
    \caption{Same as Figure \ref{fig:confronto320} for tile 3\_11.}
    \label{fig:confronto311}
\end{figure}

Figures \ref{fig:320_TRI_YMCA} and \ref{fig:311_TRI_YMCA} shows the recovered SFH for tiles 3\_20 and 3\_11, respectively, highlighting the differences introduced by the two methods. The corresponding residuals are displayed in Figures \ref{fig:confronto320} and \ref{fig:confronto311}, where the TRILEGAL-based model clearly introduces systematically larger discrepancies, especially for $g-i$ between 2 and 3 mag. The primary source of these discrepancies stems from the difficulty in accurately modeling the foreground contamination with synthetic CMDs. While TRILEGAL provides a sophisticated model for Galactic stellar populations, it lacks the ability to fully replicate the specific MW stars distribution present in our fields. In contrast, the YMCA tile directly samples the foreground contamination in a manner that is inherently consistent with the data. The residuals around the RC
region in tile 3\_20 are likely attibutable to significant theoretical modeling uncertainties for advanced stellar evolutionary phases, particularly concerning the intricate interplay of poorly understood physical processes (e.g.: late-phase stellar winds, convection, mass loss).

\section{CMDs of individual tiles}

\begin{sidewaysfigure}
  \centering
  \vspace*{100mm}
  \includegraphics[width=0.82\paperheight,trim={50 0 40 0mm}, keepaspectratio, clip]{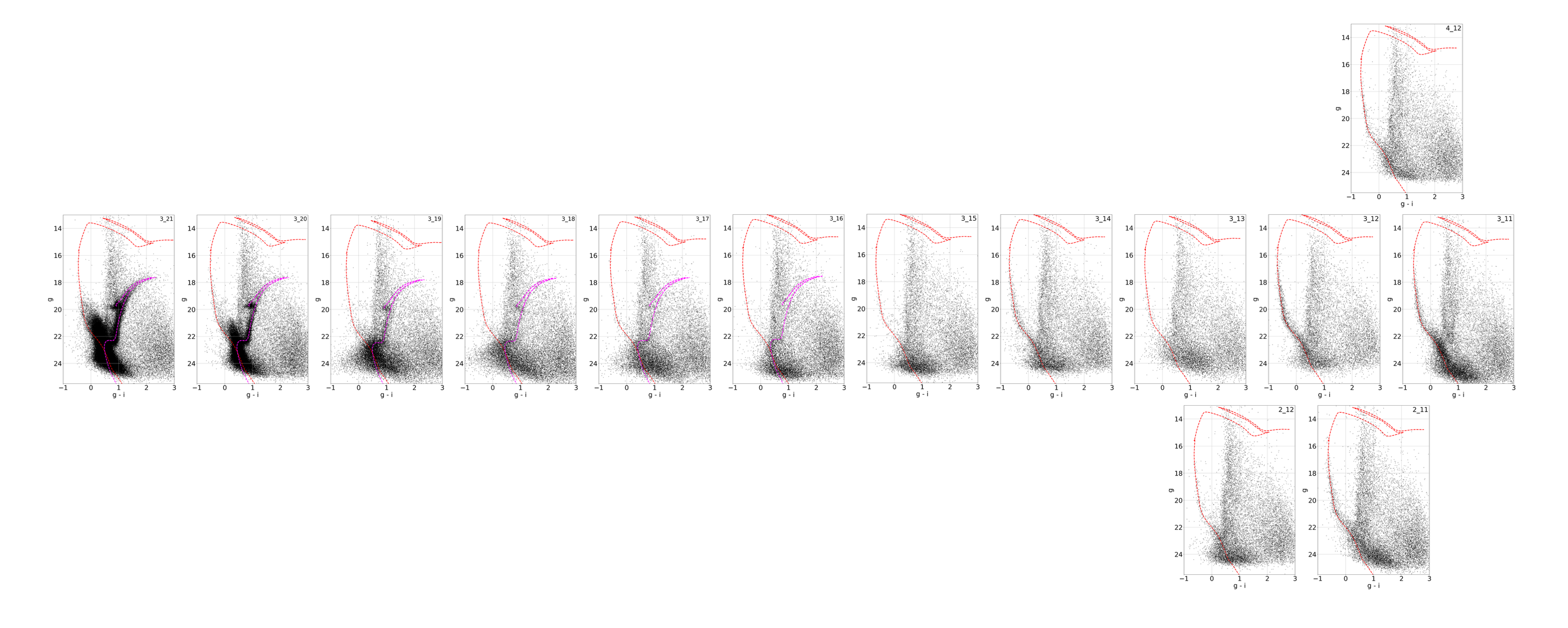}

  \caption{CMDs of the analyzed regions. We overplotted a reference isochrone representing an age of 30 Myr and [Fe/H] = -0.5 dex (red dashed line), and, from tile 3\_16 onward,  an old isochrone representing an age of 8 Gyr and [Fe/H] = -1 dex (magenta dashed line). The distance modulus and extinction applied to the isochrones are those listed in Table \ref{tab:parameters} for the corresponding region.}
  \label{fig:CMDs_tiles}
\end{sidewaysfigure}

In Figure \ref{fig:CMDs_tiles}, we present the CMDs of all the tiles included in our analysis. These deep CMDs offer an unprecedented view of the stellar populations across the Magellanic Bridge, covering the most extensive area hitherto observed. To facilitate spatial interpretation, the CMDs are arranged to reproduce the sky distribution of the tiles. A fiducial main-sequence ridge line (red dashed) is overplotted as a reference, and an old isochrone (magenta dashed line) is added in the tiles closer to the LMC. The number of stars contributing to each CMD is reported in Table \ref{tab:parameters}.

\label{sec:all_CMDs}
\clearpage


\end{document}